\documentclass[useAMS,usenatbib]{mn2e}
\usepackage{color}

\usepackage{ulem}
\usepackage{lineno}
\usepackage{longtable}
\usepackage{supertabular}
\usepackage{multicol}
\usepackage{lipsum}

\newcommand\ergs{erg~s$^{-1}$}
\newcommand\ergcms{erg~cm$^{-2}$~s$^{-1}$}

\newcommand\nodata{ ~$\cdots$~ }

\def\ergs{erg~s$^{-1}$}

\def\ergcms{erg~cm$^{-2}$~s$^{-1}$}

\usepackage{graphicx}
\usepackage{float}

\def\aap{A\&A}
\def\apj{ApJ}
\def\mnras{MNRAS}
\def\apjl{ApJL}

\def\nat{Nature}
\def\araa{ARA\&A}

\def\pasj{PASJ}
\def\procspie{Proc. SPIE}
\def\ssr{SSRv}
\def\aapr{A\&AR}
\def\nar{NewAR}

\title[MAXI~J1535-571]{\textit{Swift} Observations of the bright uncatalogued X-ray transient MAXI~J1535-571}

\author[Tao et al.]{Lian Tao,$^{1}$\thanks{E-mail: lian.taotao@gmail.com}
YuPeng Chen,$^{1}$
Can G\"{U}NG\"{O}R,$^{1}$
Yue Huang,$^{1}$
FangJun Lu,$^{1}$
\newauthor JinLu Qu,$^{1}$
LiMing Song,$^{1}$
Liang Zhang,$^{1}$
Shu Zhang$^{1}$
and ShuangNan Zhang$^{1,2,3}$ \\
$^{1}$Key Laboratory of Particle Astrophysics, Institute of High Energy Physics, Chinese Academy of Sciences, Beijing 100049, China\\
$^{2}$National Astronomical Observatories, Chinese Academy of Sciences, Beijing 100012, China\\
$^{3}$University of the Chinese Academy of Sciences, Beijing, China}

\begin{document}
\date{Accepted 2018 August 03. Received 2018 July 07; in original form 2017 November 13}

\pagerange{\pageref{firstpage}--\pageref{lastpage}} \pubyear{2017}

\maketitle

\label{firstpage}

\begin{abstract}

The black hole candidate MAXI~J1535-571 is a recently discovered X-ray transient. We report on the monitoring observations of \textit{Swift} Gamma-Ray Burst Mission during outburst. The source transits from the hard state to the intermediate state, and reaches the soft state; near the end of the outburst, it returns to the hard state passing through the low intermediate state, following a typical Q-shaped loop in the hardness-intensity diagram. During the high intermediate state, detailed spectral analyses using a multi-temperature disk model reveal that the disk luminosity is flatter than the inner disk temperature to the fourth power, and the disk temperature profile varies as the disk radius raised to the power of $-0.5$, consistent with the behavior of a slim disk, implying that the disk structure has been modified due to the high luminosity of $\sim 10^{39}$~\ergs. Meanwhile, the column density increases with the Eddington ratio, suggesting that the outflow matters are driven by radiation pressure due to the high accretion rate.

\end{abstract}


\section{Introduction}
\label{sec:intro}
A Low-mass X-ray binary system with a black hole compact object usually accretes matter from a companion star via Roche-lobe overflow and demonstrates a transient behavior. Driven by instabilities in the accretion disk \citep{Lasota2001}, it may undergo an outburst after a long quiescent period, accordingly, its luminosity can vary by many orders of magnitude. The spectra in the outburst are primarily contributed by two components, i.e., a standard thin accretion disk \citep{Shakura1973} and a corona in a power-law form. In a typical outburst, the source usually undergo several different spectral states \citep[e.g.,][]{Fender2004, Remillard2006, Done2007}. At the initial onset phase of an outburst, it is in a hard state with the spectrum dominated by a hard power-law component, and a steady jet may be present in this period. As the accretion rate increases, the source goes through an intermediate state and then reaches a soft state. In the soft state, the accretion disk dominates the emission and the power law component becomes much softer, meanwhile, the jet quenches. The source stays in soft state until the late fading stage of the outburst. Then it returns to an intermediate state with an intensity lower than the one in the previous intermediate state. At the end of an outburst, the source turns back to the hard state. Thus, through the outburst, a black hole transient typically follows an counterclockwise Q-shaped loop in the hardness-intensity diagram (HID) \citep[e.g.,][]{Belloni2005,Belloni2016}.

In some giant outbursts, the luminosity can approach to or exceed the classical Eddington limit \citep[e.g.,][]{Revnivtsev2002,Tomsick2005,Motta2017}. The accretion disk can no longer keep its thin structure and becomes geometrically thick \citep[slim disk,][]{Abramowicz1988}, in the meanwhile, advection becomes the dominant process. Photons are trapped in the disk and the emission is thus inefficient. The disk luminosity and inner disk temperature relationship is then expected to be flatter than that of the standard thin disk \citep{Watarai2000}.

In this study, we report on a giant outburst of a new X-ray transient, MAXI~J1535-571, which exhibits clear spectral evolution during the outburst. The source was serendipitously discovered with the \textit{Monitor of All-sky X-ray Image} \citep[\textit{MAXI};][]{Matsuoka2009} Gas Slit Camera (GSC) and the \textit{Neil Gehrels Swift Observatory} (\textit{Swift}) Burst Alert Telescope \citep[BAT;][]{Barthelmy2004} when it went into outburst on 2017 September 2nd \citep{ATel10699,GCN21788}. Using \textit{Swift}/X-ray Telescope \citep[XRT;][]{Burrows2005} Photon Counting (PC) mode data with the exposure time of 3279~s, the corrected X-ray position of the source was reported to be R.A.=$15^{\rm h}35^{\rm m}19.\!^{\rm s}73$, Decl.=-$57^{\circ}13^{\prime}48.\!^{\prime\prime}1$ (J2000.0) \citep{ATel10700}. The X-ray flux in the 4.0--10.0~keV band was about 34 mCrab \citep{ATel10699}, and its \textit{Swift}/XRT spectrum was well represented by an absorbed power-law model with a neutral hydrogen column density ($N_{\rm H}$) of $3.6\times 10^{22}~\rm atoms~cm^{-2}$ and a photon index of 1.53, indicating the source was in hard state \citep{ATel10700}. 

Follow-up X-ray observations of MAXI~J1535-571 were performed. Its \textit{MAXI}/GSC spectrum obtained on September 5 also agreed with an absorbed power-law model, but the photon index was about 1.84, softer than that obtained on September 2nd \citep{ATel10708}. $N_{\rm H}$ seemed to be stable ($3.4\times 10^{22}~\rm cm^{-2}$), but was larger than the line-of-sight Galactic column density, $N_{\rm H}=1.5\times 10^{22}~\rm cm^{-2}$ \citep{Kalberla2005}, suggesting that the accretion disk or the dense interstellar gas might obscure the source.

The source continued to brighten and soften on September 10-11, and the \textit{Swift}/BAT count rate in the 15--50 keV band started to decrease, indicating that the source might undergo a hard to soft state transition \citep{ATel10729,ATel10731,ATel10733}. In the following \textit{MAXI}/GSC observation taken on September 20, the source intensity in the 2--20~keV band increased to 5~Crab, and the photon index reached to the value of 2.9. The thermal component also appeared in the spectrum; the inner disk temperature was about 1.5~keV throughout the fitting with a multi-temperature disk blackbody model ({\tt diskbb}\footnote{https://heasarc.gsfc.nasa.gov/xanadu/xspec/manual/node159.html}) \citep{ATel10761}. Later works performed by the \textit{MAXI} team obtain a more detailed look at the state transition of the source \citep{Nakahira2018}. They argued that two counterclockwise circles appeared on the HID. The source stayed at the soft state until 2018 April 30; then it showed spectral hardening in \textit{Swift}/XRT observations and went back to the hard state \citep{Russell2018}.

The \textit{Swift}/XRT power density spectrum of September 11 showed a strong low-frequency quasi-periodic oscillation (QPO) with the first and second harmonics at 1.87~Hz and 3.87~Hz, respectively \citep{ATel10734}. The QPOs also were detected by \textit{Neutron star Interior Composition ExploreR} \citep[\textit{NICER};][]{Gendreau2012} between September 12 and 13 \citep{ATel10768}. The frequency of the first harmonic drifted between 1.9 Hz and 2.8 Hz, with the fractional root-mean-squared (rms) amplitude to be 6\%, while the frequency of the second harmonic drifted between 3.8 Hz and 5.6 Hz, with the fractional rms of 5\% \citep{ATel10768}. Benefiting from the large effective area of \textit{Insight}-HXMT, QPOs were also definitely detected in the energy band up to 100~keV \citep{Huang2018}. Recently, some works reported the detailed evolution of the QPO parameters \citep{Mereminskiy2018,Shang2018}.  

%iron line
\textit{Nuclear Spectroscopic Telescope Array} \citep[\textit{NuSTAR;}][]{Harrison2013} detected a broad Fe $K_{\alpha}$ line and a Compton hump with the peak energy of 30 keV on September 7 \citep{Xu2018}. Using a model combining a multi-temperature thermal component, a lamppost reflection model \citep[{\tt relxilllpCp};][]{Dauser2014,Garca2014} and an unblurred reflection model \citep[{\tt xillverCp};][]{Garca2010}, they found the BH spin as $a>0.84$, the inner disk radius as $r_{in}~<~2.01~r_{\rm ISCO}$\footnote{$r_{\rm ISCO}$ is the innermost stable circular orbit.}, and the inclination angle of the inner disk $i(^{\circ})=57^{+1}_{-2}$. They also reported the existence of a broad iron emission line in the \textit{NICER} spectra taken on September 13. Using the model of {\tt relxill} \citep{Dauser2014,Garca2014} which is a combination of a multi-temperature thermal component and a relativistic reflection, \citet{ATel10768} obtained a consistent BH spin, $a=0.88^{+0.1}_{-0.2}$, but much lower inclination, $i(^{\circ})=27^{+1}_{-5}$. \citet{Miller2018} recently reanalyzed the \textit{NICER} data and obtained a near-maximal spin of 0.994. Moreover, they found that the accretion disk is likely to be warped and the inclination of the inner disk is $i(^{\circ})=67 \pm 8$.

%optical and IR
Multiwavelength observations of this source were also performed. The optical counterpart was identified by the 0.61m B\&C Telescope at Mt. John Observatory. The magnitude in the SDSS $i$-band was 21$^{m}$.79, while no source was detected in the SDSS $g$ and $r$ bands, suggesting that the source might be a low mass X-ray binary (LMXB) \citep{ATel10702}. The near-infrared (NIR) counterpart was identified in the following observations taken on September 5, by the Small and Moderate Aperture Research Telescope System (SMARTS) 1.3m telescope at Cerro Tololo Inter-American Observatory (CTIO) \citep{ATel10716}. \textit{Gemini} observed the source on October 2, and found that the source had faded by about 2$^m$.2 in the $H$-band. A Brackett-gamma emission line appeared in the K-band spectrum with the line centroid at 21667 \AA~with the FWHM of $79.0$ \AA~($\sim 1090 \pm 60$ km/s) \citep{ATel10816}. 

%radio
The radio counterpart of MAXI~J1535-571 was discovered by the Australia Telescope Compact Array (ATCA) on September 5 \citep{ATel10711}. The flux densities at 5.5 GHz and 9.0 GHz had a power-law form with an index of $0.09 \pm 0.03$, consistent with emission from a compact jet. The Atacama Large Millimetre/Sub-Millimetre Array (ALMA) and the ATCA then observed the radio source on September 11 and 12, respectively \citep{ATel10745}. A spectral break between the 140--230 GHz was found, which means that the optically thick jet changed to optically thin at this frequency range.

%the nature of the source
Inferred from the X-ray and multiwavlength observations, MAXI~J1535-571 differs from a neutron star. Its radio luminosity at 5 GHz is well above the limit of all known neutron star binaries, but consistent with the values of BH binaries \citep{ATel10711}. Moreover, its \textit{MAXI}/GSC light curve shows rapid X-ray variability with no obvious pulsation \citep{ATel10708} and is lack of X-ray bursts. Moreover, its unabsorbed flux of the 1--60 keV in the onset of the outburst is $\sim 3.1\times 10^{-8}$~\ergcms; assuming a typical distance of Galactic black holes (BHs) of 8~kpc, the luminosity is $\sim 2.4\times 10^{38}$~\ergs~and violates the classical Eddington limit of a neutron star \citep{ATel10708}. However, the dynamical mass constraint is not available yet, thus we prefer to mention that it is still a BH candidate, instead of firmly conclude its BH nature.

In this paper, we particularly use \textit{Swift} monitoring observations (Section~\ref{sec:obs}) to investigate the spectral evolution and properties of the recently discovered BH candidate, MAXI~J1535-571 (Section~\ref{sec:res}), and then discuss the results in Section~\ref{sec:dis}.

%http://www.astronomerstelegram.org/?read=10761

%%%%%%%%%%%%%%%%%%%%%%%%%%%%%%%%%%%%%%%%%%%%%%%%%%%%%%%%%%%%%%%%%%%%%%%%%%
\begin{figure*}
\centering
\includegraphics[width=0.98\textwidth]{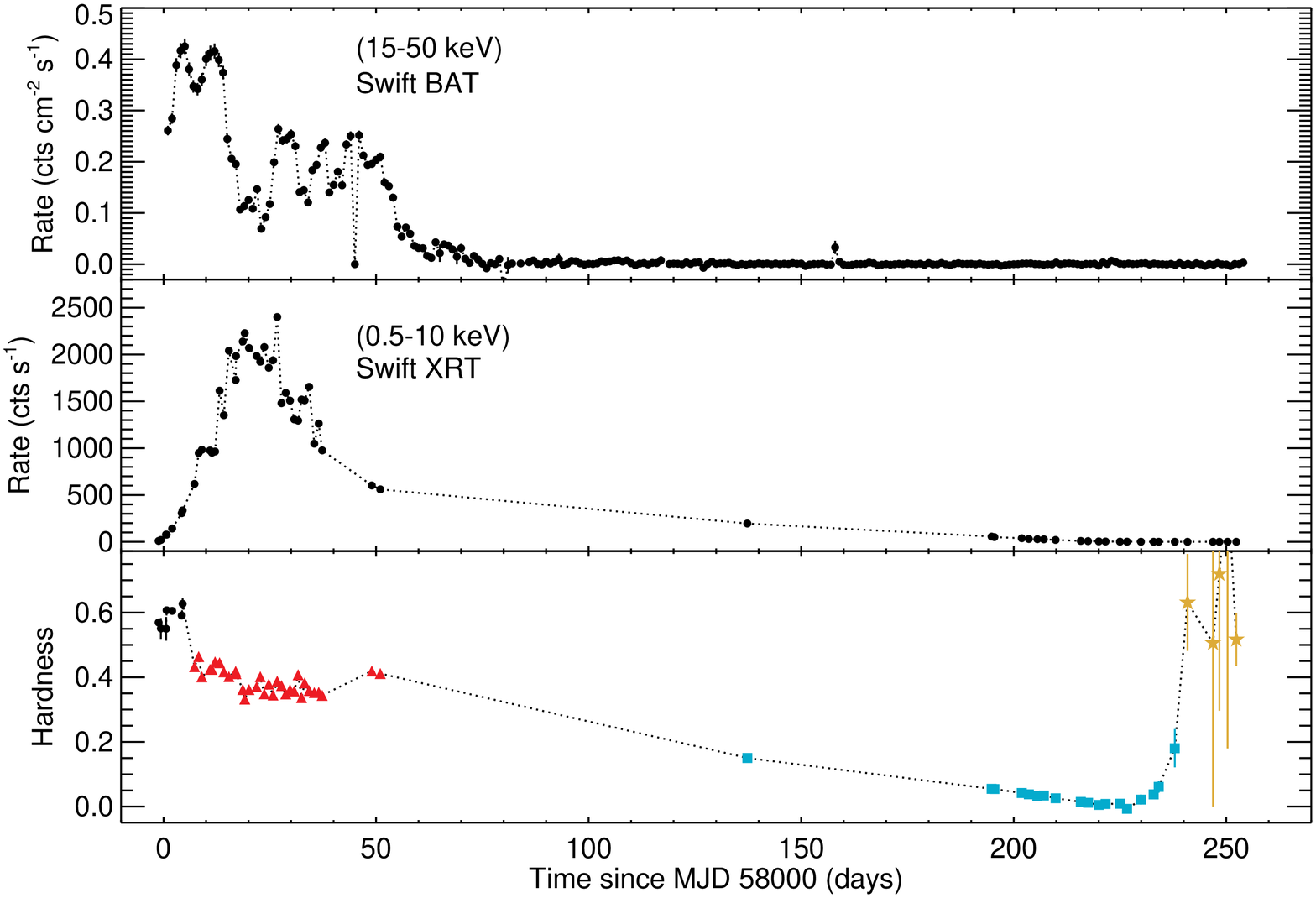} \\
\caption{The 15--50 keV \textit{Swift}/BAT count rate (top panel), the 0.5--10 keV \textit{Swift}/XRT count rate (middle panel), and the hardness ratio of the \textit{Swift}/XRT hard band (4.0--10 keV) to the soft band (0.5--4.0 keV; the bottom panel). Black dots, red triangles, blue squares and golden stars in the bottom panel indicate four distinguishable groups in the hardness ratio, respectively.
\label{fig:cts}}
\end{figure*}
%%%%%%%%%%%%%%%%%%%%%%%%%%%%%%%%%%%%%%%%%%%%%%%%%%%%%%%%%%%%%%%%%%%%%%%%%%

\section{OBSERVATIONS and Data reduction}
\label{sec:obs}
Monitoring of MAXI J1535-571 with \textit{Swift} commenced on 2017 September 2. In order to study the outburst properties of the source, we search for all of \textit{Swift}/XRT observations until 2018 May 14th. Due to its extreme brightness, except for the late fading stage, we only use the data taken in windowed timing (WT) mode. In the late fading stage, the data taken in photon counting (PC) mode are used instead as the source is very weak. The properties of the observations are listed in Table~\ref{tab:mod}.

%%%%%%%%%%%%%%%%%%%%%%%%%%%%%%%%%%%%%%%%%%%%%%%%%%%%%%%%%%%%%%%%%%%%%%%%%%
\begin{table*}
\begin{minipage}{160mm}
\centering
\caption{General properties of \textit{Swift}/XRT Observations sampled in this paper. Most of the \textit{Swift}/XRT data are taken in windowed timing mode except some observations in the late fading stage. For these observations in the late stage, we use the data obtained in photon counting mode instead. Hard -- hard state; HIM -- hard intermediate state; SIM -- soft intermediate state; Soft -- soft state.}
\label{tab:obs}
\begin{tabular}{lcclllll}
\hline
ObsID  & Observed date & Observed date & Exposure  & Count rate$^{b}$  & Count rate$^{c}$ &  Pixels$^{d}$ & State\\
  & (day) & (MJD) & (s) & (cts~s$^{-1}$) & (cts~s$^{-1}$) &  & \\
\hline
770431000 & 2017-09-02~20:07:41 & 57998.84 & 1615.86  & $9.79 \pm 0.08$ & $9.31\pm 0.08$ & 0 & Hard \\
770502000 & 2017-09-03~09:02:43 & 57999.38 & 75.03  & $20.8 \pm 0.5$ & $ 19.9\pm 0.6 $  &  0 & Hard\\
770656000 & 2017-09-04~14:05:48 & 58000.59 & 15.60  & $74 \pm 2 $ & $ 77 \pm 2$   & 0  & Hard\\
770656001 & 2017-09-04~17:12:35 & 58000.72 & 988.80  & $71.1 \pm 0.3$ & $ 76.1\pm 0.3$   & 0 & Hard\\
10264002 & 2017-09-06~00:41:58 & 58002.03 & 449.23  & $139.1 \pm 0.6$ & $ 142.0\pm 0.8$  & 2  & Hard\\
10264003 & 2017-09-08~06:40:51 & 58004.28 & 4948.53  & $221.7 \pm 0.2$ & $306.1 \pm 0.5$  & 3  & Hard\\
771371000 & 2017-09-08~11:39:18 & 58004.49 & 69.12  & $236 \pm 2$ &  $ 335\pm 4$  & 4 & Hard\\
10264004 & 2017-09-11~06:34:43 & 58007.27 & 1089.61  & $510.4 \pm 0.7$ & $ 618\pm 2 $  &  7 & HIM \\
10264005 & 2017-09-12~06:19:10 & 58008.26 & 899.61  & $543.7 \pm 0.8$ & $ 948\pm 4$  & 9  & HIM\\
10264007 & 2017-09-13~00:11:05 & 58009.01 & 889.45  & $661.2 \pm 0.9$ & $ 983\pm 4$  & 9  & HIM\\
10264006 & 2017-09-14~22:19:59 & 58010.93 & 1074.61  & $640.5 \pm 0.8 $ & $ 975\pm 3 $  & 9  & HIM\\
10264008 & 2017-09-15~09:19:30 & 58011.39 & 984.64  & $643.2 \pm 0.8$ & $ 951\pm 3$  & 8  & HIM\\
10264009 & 2017-09-16~02:47:14 & 58012.12 & 1059.79  & $644.5 \pm 0.8$ & $ 964\pm 3$  & 9  & HIM\\
88245001 & 2017-09-17~04:17:53 & 58013.18 & 2223.08  & $690.1 \pm 0.6$ & $ 1613\pm 4$  & 11  & HIM\\
10264010 & 2017-09-18~04:13:31 & 58014.18 & 1104.61  & $783.5 \pm 0.8$ & $ 1352\pm 5 $  & 10  & HIM\\
10264011 & 2017-09-19~08:56:26 & 58015.37 & 989.39  & $29.5 \pm 0.2$ & $ 2041\pm 13$  & 0$^e$  & SIM\\
88245002 & 2017-09-20~23:13:58 & 58016.97 & 954.61  & $933.6 \pm 1.0$ & $ 1728 \pm 6$  & 11  & SIM\\
88245003 & 2017-09-21~00:48:59 & 58017.03 & 1014.19  & $1012.0 \pm 1.0$ & $1984 \pm 8$ & 13  & SIM\\
10264012 & 2017-09-22~15:17:53 & 58018.64 & 1189.21  & $998.4 \pm 0.9$ & $ 2139\pm 7$  & 13 & SIM\\
10264013 & 2017-09-23~02:23:05 & 58019.10 & 1009.60  & $1066.7 \pm 1.0$  & $2228 \pm 8$  &13  & SIM\\
10264014 & 2017-09-24~02:29:57 & 58020.10 & 599.21  & $836.4 \pm 1.2 $  & $2070 \pm 10$ &13   & SIM\\
10264015 & 2017-09-25~21:30:52 & 58021.90 & 899.49  & $972.9 \pm 1.0$  & $1984 \pm 8$  &13  & SIM\\
10264016 & 2017-09-26~18:15:01 & 58022.76 & 909.22  & $942.4 \pm 1.0$  & $ 1923\pm 8$  &13  & SIM\\
10264017 & 2017-09-27~16:40:13 & 58023.70 & 859.22  & $875.5 \pm 1.0$  & $ 2079\pm 8$ & 11  & SIM\\
10264018 & 2017-09-28~18:04:31 & 58024.75 & 1044.61  & $880.3 \pm 0.9$  & $ 1858\pm 7$  & 12 & SIM\\
10264019 & 2017-09-29~18:00:51 & 58025.75 & 964.65  & $778.8 \pm 0.9$  & $ 1938\pm 7$  &12  & SIM\\
10264020 & 2017-09-30~18:01:05 & 58026.75 & 889.62  & $899.3 \pm 1.0$  & $ 2401\pm 10$ & 13  & SIM\\
10264021 & 2017-10-01~17:51:20 & 58027.74 & 874.62  & $742.2 \pm 0.9$  & $ 1480 \pm 6$ & 12  & SIM\\
10264022 & 2017-10-02~17:48:25 & 58028.74 & 1109.61  & $842.3 \pm 0.9$  & $ 1590\pm 6 $  &12  & SIM\\
10264023 & 2017-10-03~17:46:07 & 58029.74 & 1004.61  & $878.5 \pm 0.9$  & $ 1507\pm 6$  & 11 & SIM\\
10264024 & 2017-10-04~17:44:51 & 58030.74 & 964.61  & $358.1 \pm 0.6$  &  $ 1307\pm 5$ & 11 & SIM\\
10264025 & 2017-10-05~15:55:38 & 58031.66 & 604.61  & $754.5 \pm 1.1$  &  $ 1294\pm 6$ & 10  & SIM\\
10264026 & 2017-10-06~11:08:37 & 58032.47 & 1154.61  & $867.0 \pm 0.9$  & $ 1519\pm 5$  & 11 & SIM\\
10264027 & 2017-10-07~04:37:47 & 58033.19 & 1024.61  & $429.3 \pm 0.6$  & $ 1510\pm 5$  & 11 & SIM\\
10264028 & 2017-10-08~05:57:40 & 58034.25 & 974.61  & $814.6 \pm 0.9$  &  $ 1655\pm 7$ & 12 & SIM\\
10264029 & 2017-10-09~10:54:12 & 58035.46 & 1059.61  & $560.1 \pm 0.7$  &  $1048\pm 4$ & 10 & SIM\\
10264030 & 2017-10-10~12:07:37 & 58036.51 & 994.06  & $705.6 \pm 0.8$  &  $1263\pm 5$ & 10 & SIM\\
10264031 & 2017-10-11~07:35:39 & 58037.32 & 854.62  & $651.2 \pm 0.9$ & $976\pm 4$ & 9  & SIM\\
88245004 & 2017-10-22~23:42:52 & 58048.99 & 2088.03 & $434.1 \pm 0.5$ & $602\pm 2$  & 6  & SIM\\
88246001 & 2017-10-24~23:31:53 & 58050.98 & 1528.20 & $392.9 \pm 0.5$ & $560\pm 2$  & 6  & SIM\\
10491002 & 2018-01-19~08:08:13 & 58137.34 & 1119.61 & $171.7 \pm 0.4$ & $195.2\pm 0.7$  & 3  & Soft \\
10264032 & 2018-03-17~19:13:28 & 58194.80 & 279.62 & $34.1 \pm 0.3$ & $55.7\pm 0.6 $  & 0  & Soft\\
10264033 & 2018-03-18~10:47:57 & 58195.45 & 1370.35 & $48.1 \pm 0.19$ & $50.1\pm 0.2$  & 0  & Soft\\
10264034 & 2018-03-24~21:16:57 & 58201.89 & 1859.22 & $28.93 \pm 0.12$ & $37.78\pm 0.18$  & 0  & Soft\\
10264035 & 2018-03-26~13:28:47 & 58203.56 & 1409.22 & $19.56 \pm 0.12$ & $30.4\pm 0.2$  & 0  & Soft\\
10264036 & 2018-03-28~13:23:57 & 58205.56 & 2015.28 & $25.39 \pm 0.11$ & $27.96\pm 0.13$  & 0  & Soft\\
10264037 & 2018-03-30~02:03:57 & 58207.09 & 1910.28 & $23.88 \pm 0.11$ & $26.68\pm 0.14$  & 0  & Soft\\
10264038 & 2018-04-01~20:30:56 & 58209.86 & 509.46 & $20.2 \pm 0.2$ & $19.8\pm 0.2$  & 0  & Soft\\
10264039 & 2018-04-07~18:47:57 & 58215.79 & 2244.15 & $8.09 \pm 0.06$ & $9.21\pm 0.08$  & 0  & Soft\\
10264040 & 2018-04-09~11:41:57 & 58217.49 & 2131.63 & $7.36 \pm 0.06$ & $7.00\pm 0.07$  & 0  & Soft\\
10264041 & 2018-04-12~00:32:57 & 58220.03 & 1949.17 & $4.23 \pm 0.05$ & $4.32\pm 0.06$  & 0  & Soft\\
10264042 & 2018-04-13~14:35:57 & 58221.61 & 1719.23 & $3.08 \pm 0.04$ & $3.13\pm 0.06$  & 0  & Soft\\
10264043 & 2018-04-16~23:58:57 & 58225.00 & 1115.41 & $2.14 \pm 0.04$ & $1.53\pm 0.05$  & 0  & Soft\\
10264044 & 2018-04-18~15:51:57 & 58226.66 & 249.45 & $1.47 \pm 0.08$ & $1.21\pm 0.09$  & 0  & Soft\\
10264045 & 2018-04-21~22:21:57 & 58229.94 & 1063.92 & $0.82 \pm 0.03$ & $0.53\pm 0.04$  & 0  & Soft\\
10264046$^a$ & 2018-04-24~22:04:57 & 58232.92 & 1703.15 & $0.180 \pm 0.010$ & $0.195\pm 0.011$  & 0  & IM\\
10264047$^a$ & 2018-04-26~02:32:56 & 58234.11 & 1797.36 & $0.118 \pm 0.008$ & $0.129\pm 0.009$  & 0  & IM\\
\hline
\end{tabular}
\end{minipage}
\end{table*}
%%%%%%%%%%%%%%%%%%%%%%%%%%%%%%%%%%%%%%%%%%%%%%%%%%%%%%%%%%%%%%%%%%%%%%%%%%

%%%%%%%%%%%%%%%%%%%%%%%%%%%%%%%%%%%%%%%%%%%%%%%%%%%%%%%%%%%%%%%%%%%%%%%%%%
\begin{table*}
\begin{minipage}{160mm}
\centering
\contcaption{}
%\label{tab:obs}
\begin{tabular}{lcclllll}
\hline
ObsID  & Observed date & Observed date & Exposure & Count rate$^{b}$ & Count rate$^{c}$ &  Pixels$^{d}$ & State \\
  & (day) & (MJD) & (s) & (cts~s$^{-1}$) & (cts~s$^{-1}$) & & \\
\hline
10264048$^a$ & 2018-04-29~21:21:57 & 58237.89 & 1771.22 & $0.042 \pm 0.005$ & $0.049\pm 0.006$  & 0  & Hard \\
10264049$^a$ & 2018-05-02~21:23:57 & 58240.90 & 1854.74 & $0.041 \pm 0.005$ & $0.047\pm 0.005$  & 0  & Hard\\
10264050$^a$ & 2018-05-08~20:48:58 & 58246.87 & 1790.09 & $0.005 \pm 0.003$ & $0.005\pm 0.003$  & 0  & Hard\\
10264051$^a$ & 2018-05-10~09:17:57 & 58248.39 & 1903.64 & $0.006 \pm 0.002$ & $0.009\pm 0.003$  & 0  & Hard\\
10264052$^a$ & 2018-05-12~07:17:56 & 58250.31 & 1003.63 & $0.006 \pm 0.002$ & $0.008\pm 0.003$  & 0  & Hard\\
10264053$^a$ & 2018-05-14~08:41:38 & 58252.37 & 1470.92 & $0.125 \pm 0.009$ & $0.143\pm 0.011$  & 0  & Hard\\
\hline
\end{tabular}
\footnotesize{
\begin{flushleft} $^a$The observations are taken in photon counting mode.\\
$^b$The observed count rates, which are not corrected for bad pixels, background and pileup. \\
$^c$The count rates in the 0.5--10.0 keV band have been corrected for bad pixels, background and pileup, using the PSF correction factor in {\tt xrtmkarf}. \\
$^d$The radii of the excluded region. \\
$^e$During this observation, \textit{Swift} partly missed the source. 
\end{flushleft} }
\end{minipage}
\end{table*}
%%%%%%%%%%%%%%%%%%%%%%%%%%%%%%%%%%%%%%%%%%%%%%%%%%%%%%%%%%%%%%%%%%%%%%%%%%

%%%%%%%%%%%%%%%%%%%%%%%%%%%%%%%%%%%%%%%%%%%%%%%%%%%%%%%%%%%%%%%%%%%%%%%%%%
\begin{figure*}
\centering
\includegraphics[width=0.45\textwidth]{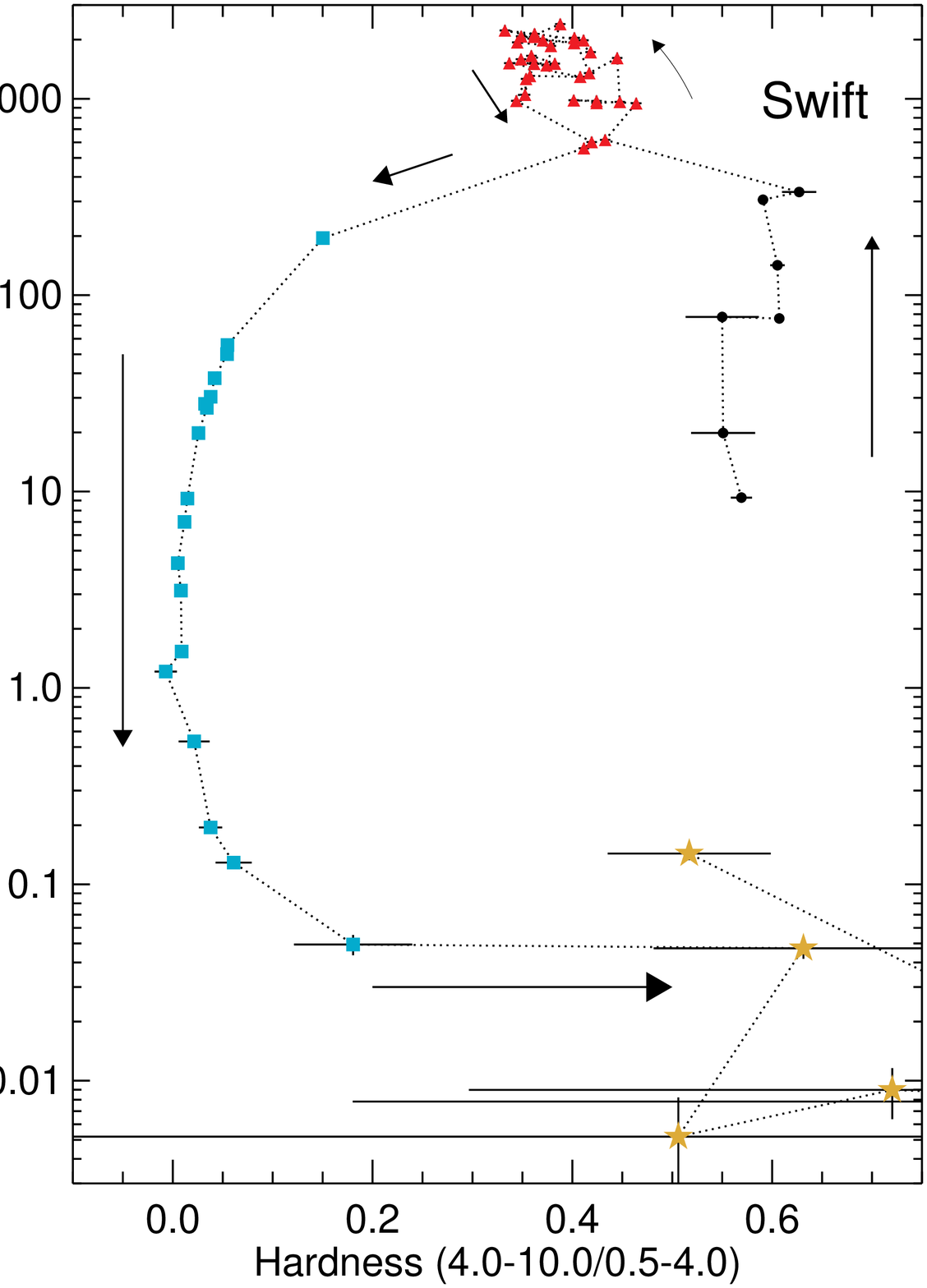}\hfill
\includegraphics[width=0.45\textwidth]{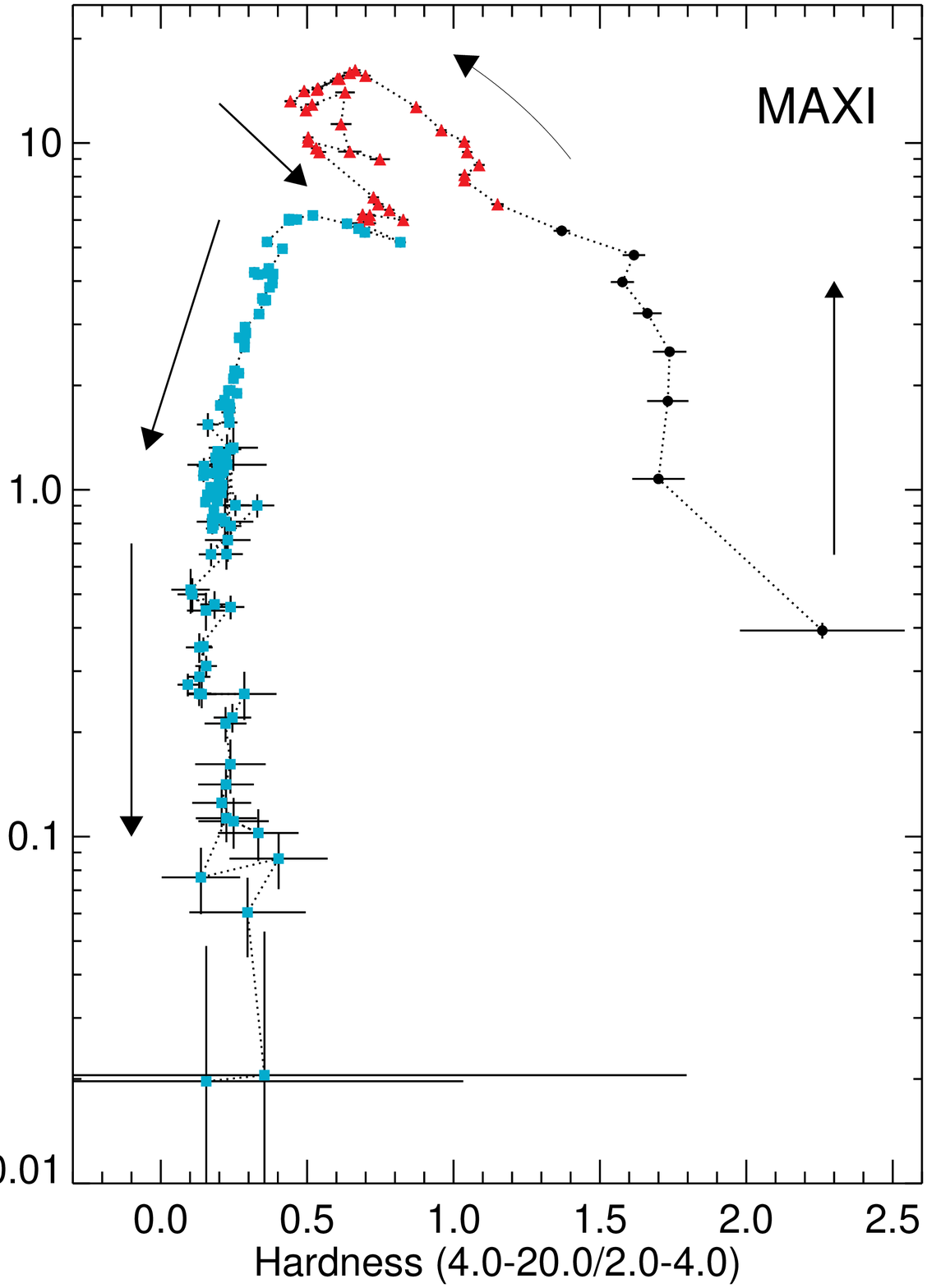}
\caption{\textit{Swift}/XRT and \textit{MAXI}/GSC hardness-intensity diagrams of the source. Left panel: \textit{Swift}/XRT HID. Intensities here are the count rates in the 0.5--10.0 keV, while hardness values are obtained using the ratio of 4--10 keV to 0.5--4.0 keV count rates. Right panel: \textit{MAXI}/GSC HID. Intensities are the count rates in the 2--20 keV, while hardness values are the ratios of 4--20 keV to 2.0--4.0 keV count rates. The symbols here are the same as Figure~\ref{fig:cts}. Arrows show the evolutionary track of the outburst.  
\label{fig:hid}}
\end{figure*}
%%%%%%%%%%%%%%%%%%%%%%%%%%%%%%%%%%%%%%%%%%%%%%%%%%%%%%%%%%%%%%%%%%%%%%%%%%

The \textit{Swift}/XRT data are reduced using XSELECT, HEASoft v6.22, with XRT CALDB version 20170501. However, most of the observations are faced with several pileups due to the high count rates ($>100$ cts~s$^{-1}$, Table~\ref{tab:obs}) of the source. We thus extract the counts by excluding the core using an annulus if pileup is present in the observation. The inner radius of the excluded region is selected so that the extracted count rate falls below 100 cts~s$^{-1}$ (method 1, Table~\ref{tab:obs}). For comparison, we also investigate the level of pileup by increasing the excluded region and seeking for the key radius where the spectra is no longer soften (method 2, see Section~\ref{sec:spec} for more details) or the fraction of grade 0 events stay close to constant (method 3) \footnote{http://www.swift.ac.uk/analysis/xrt/pileup.php}. Method 1 and 2 give a consistent result. Method 3, however, shows a trend towards a smaller excluded region, and thus is not strict as the method 1 and 2.

The outer radius of the extracted region is 20 pixels. For WT mode data, the background extraction region is an annulus centered at the source with the inner and the outer radii of 90 and 110 pixels, respectively. For PC mode data, the background extraction region is a source-free annulus centered at the source with the inner and the outer radii of 60 and 110 pixels, respectively. Using the ftool {\tt xrtmkarf} including the point-like PSF correction, ancillary response files are created and the ratios between the total fluence of the source and the fluence within the annular extraction region are estimated. The loss of counts due to pileup and bad pixels are then corrected using the fluence ratio. Finally, except for the last six observations (ObsID 10264048 -- 10264053), the extracted spectra are rebinned to contain minimum 20 counts per bin using {\tt grppha} in order to use $\chi^2$ statistics in the spectral fits below (section~\ref{sec:spec}). For ObsID 10264048 -- 10264053, there are very few photons, the spectra are thus grouped with a minimum of 1 count in each bin, and we use C statistic in the spectral fits.

\section{Results}
\label{sec:res}

\subsection{Light Curve and Hardness Ratio}
As shown in the light curves (Figure~\ref{fig:cts}) of {\textit{Swift}/BAT} (15--50 keV band)\footnote{https://swift.gsfc.nasa.gov/results/transients/weak/MAXIJ1535-571/} and XRT (0.5--10 keV), the source went through a clear evolution and was extremely variable over the duration of the outburst. The hardness ratio, defined as the ratio of the count rates in the 4.0--10 keV to 0.5--4.0 keV bands, exhibited four distinguishable groups (Figure~\ref{fig:cts}): it stayed around 0.5--0.6 at the beginning of the outburst (group A), but suddenly decreased to $\sim 0.4$ on MJD 58007 (group B), and then stayed close to that level; after a ninety-day interval that \textit{Swift} did not observe the source, the hardness ratio was less than 0.2, which is much smaller than the previous values (group C); near the end of the outburst, the hardness ratio suddenly jumped up to a high level, with a value of around 0.6, similar to that of the beginning of the outburst (group D). Different groups in the hardness ratio may indicate different spectral state. Therefore, the source may undergo several state transitions over the course of the outburst.

HIDs are powerful tools to distinguish different spectral states. We thus plotted two HIDs using \textit{Swift}/XRT observations in the energy range of 0.5--10.0 keV from MJD 57999 to MJD 58252 and the simultaneous \textit{MAXI}/GSC data \footnote{http://134.160.243.77/star\_data/J1535-572/J1535-572.html} in the energy range of 2--20 keV, and marked the four groups above with the same symbols used for the hardness ratio curve (Figure~\ref{fig:hid}). Near the end of the outburst, the source was too weak to be detected by \textit{MAXI}, thus group D is not in the \textit{MAXI} HID. Two HIDs follow a similar X-ray spectral evolution track, similar to the typical Q-shaped loop of BH binaries \citep[Figure 4 of ][]{Belloni2016}. Group A and D are in the right branches of the HIDs, which means that the source was in the hard state of raising stage and fading stage, respectively. Group C is in the left branches of the HIDs, which indicates that the source was in the soft state. Group B goes through a small counter clockwise loop, and it is referred to the intermediate state. Since several pileups are present in most of \textit{Swift}/XRT observations, the central source pixels are excluded and the final enough count rate were not enough to do detailed timing analyses. However, benefiting from pileup-free and large effective area of \textit{Insight}-HXMT, based on the timing analyses, \citet{Huang2018} argued that the source was in the intermediate state from MJD 58008 to MJD 58019, which is consistent with our results. Moreover, the spectral analyses explained in Section~\ref{sec:spec} verifies our state identification.

\subsection{Spectral Properties}
\label{sec:spec}
The XSPEC v12.9.1m software package \citep{Arnaud1996} is used to perform the spectral fits. For the absorption calculation, the absorption model ({\tt tbabs}) with \citet{Wilms2000} abundances and \citet{Verner1996} cross-sections, is implemented. The \textit{Swift}/XRT data below 1.0 keV are ignored during the spectral fits in order to exclude the low-energy spectral residuals in windowed timing mode. For PC mode, the spectra in the energy range of 0.3--10.0~keV is adopted to the spectral fits. If pileup is present in an observation, the level of pileup is investigated by increasing the excluded region. We find that the photon index ($\Gamma$) stays stable when using the radius listed in Table~\ref{tab:obs}, and the typical change in $\Gamma$ is less than 0.02 when continuing to increase the excluded region. The spectra of BH binaries are usually a composite of a thermal component, represented by a {\tt diskbb} model, and a nonthermal component, represented by a power-law model \citep{Remillard2006}. We thus experiment several models, i.e., a single power-law, a single {\tt diskbb} and a two-component model consisting of power-law and disk components, to explore the significance of the disk and power-law components. The two-component model is preferred only if the F-test probability is low ($<0.05$). However, near the end of the outburst (ObsID 10264046 -- 10264053), the source is so weak, therefore count rates are not enough to decide the best fitting model, we thus inferred it from the hardness raio (Figure~\ref{fig:cts}). For ObsID 10264046 -- 10264047, the hardness ratio started to increase and the source passed through the hard state after four days, thus the two-component model is used. For ObsID 10264048 -- 10264053, a single power-law is preferred as the source has already come back to the hard state. The models adopted are summarized in Table~\ref{tab:obs}. 

\begin{figure}
\centering
\includegraphics[width=0.48\textwidth]{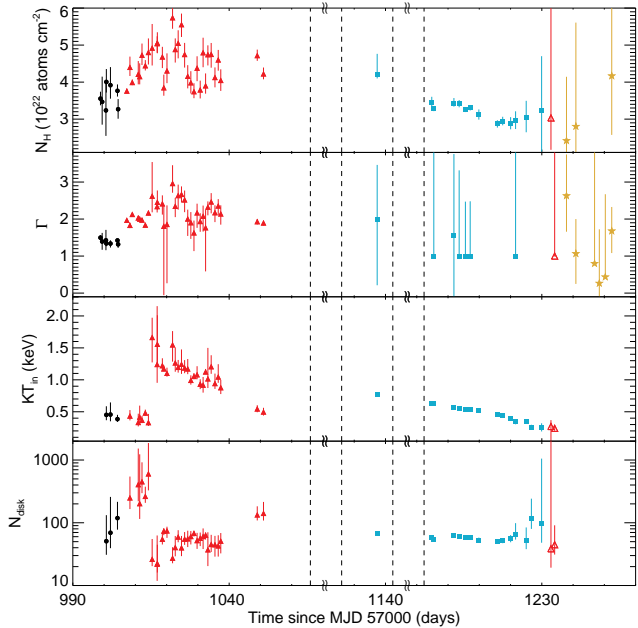} \\
\caption{Evolution of \textit{Swift}/XRT spectral parameters. The fixed parameters or the one with unconstrainable errors are not plotted in this figure. $N_{\rm H}$ is the X-ray absorption column density; $\Gamma$ is the photon index; $T_{\rm in}$ is the inner disk temperature; $N_{\rm disk} = R_{\rm in}(\rm km)*D_{10}^{-1}*\sqrt{cos(i)}$ is the square root of the normalization of the diskbb model, where $R_{\rm in}(\rm km)$ is the inner disk radius, $D_{10}$ is the source distance in units of 10 kpc, and $i$ is the angle of the disk. We marked different states using the same symbols as Figure~\ref{fig:cts}, except for ObsID 10264046 -- 10264047, which are revised as intermediate state based on spectral properties; hard state during the outburst rise (black dots), intermediate state during the outburst rise (red filled triangles), soft state (blue squares), intermediate state in the fading stage (red opening triangles), hard state in the fading stage (golden stars).
\label{fig:spe}}
\end{figure}

\begin{figure}
\centering
\includegraphics[width=0.48\textwidth]{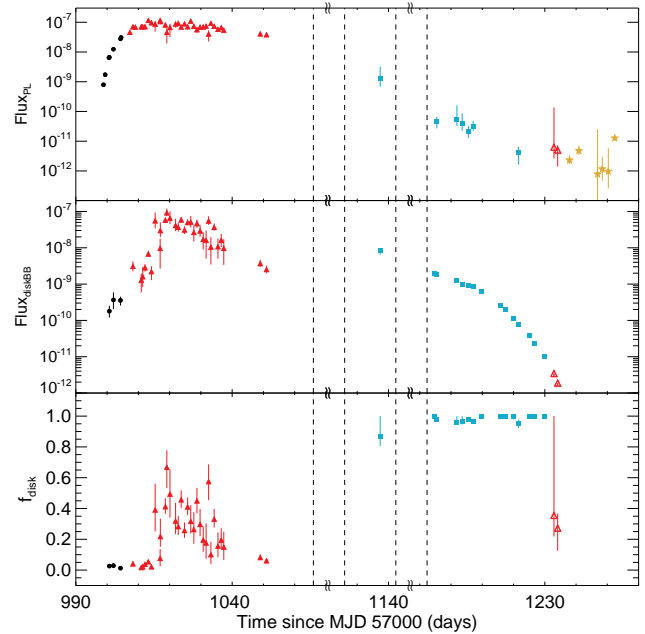} \\
\caption{Evolution of absorbed flux of the power-law and {\tt diskbb} models in the energy range of the 0.5--10 keV, and the flux ratio of the disk component. Flux here is in units of \ergcms. Different states are marked using the same colored symbols as Figure~\ref{fig:spe}.
\label{fig:flu}}
\end{figure}

The X-ray absorption column density $N_{\rm H}$, the photon index $\Gamma$, the inner disk temperature $T_{\rm in}$, the square root of the normalization of the diskbb model ($N_{\rm disk} = R_{\rm in}(\rm km)*D_{10}^{-1}*\sqrt{cos(i)}$) are shown in Figure~\ref{fig:spe}. The absorbed flux of the power-law and disk components in the 0.5--10.0 keV band, and the evolution of the ratios of the disk component flux to the total flux are shown in Figure~\ref{fig:flu}. Parameter errors around the peak of outburst are larger than that in the raising stage and early fading stage, which seems to be in strong contrast with the expected uncertainty in statistics. However, we should note that the `detected' counts are similar in these stages due to the pileup-excluding criteria ($<100$~cts~s$^{-1}$), and the PSF correction factor in {\tt xrtmkarf} are needed to correct the `real' counts. Therefore, although the `detected' statistical errors in these stages are similar, the actual uncertainties around the outburst peak became much larger due to a large PSF correction factor.

%%%%%%%%%%%%%%%%%%%%%%%%%%%%%%%%%%%%%%%%%%%%%%%%%%%%%%%%%%%%%%%%%%%%%%%%%%
%%%%%%%%%%%%%%%%%%%%%%%%%%%%%%%%%%%%%%%%%%%%%%%%%%%%%%%%%%%%%%%%%%%%%%%%%%

%%%%%%%%%%%%%%%%%%%%%%%%%%%%%%%%%%%%%%%%%%%%%%%%%%%%%%%%%%%%%%%%%%%%%%%%%%
%%%%%%%%%%%%%%%%%%%%%%%%%%%%%%%%%%%%%%%%%%%%%%%%%%%%%%%%%%%%%%%%%%%%%%%%%%

Compared with the Galactic absorption column density \citep[$N_{\rm H}=1.5\times 10^{22}~\rm cm^{-2}$, ][]{Kalberla2005}, $N_{\rm H}$ show an obvious excess in the whole outburst, indicating that some intrinsic materials, such as the accretion disk, interstellar gas and outflow matter, may obscure the source. Moreover, $N_{\rm H}$ evolves along the outburst and reaches its peak value around the outburst maxima. 

At the beginning of the outburst (MJD 57999 -- 58004), the spectra were hard ($\Gamma$ $\sim$1.5), and the disk component was cool ($T_{\rm in}<$ 0.5~keV) and weak (disk fraction $f_{\rm disk} <$ 5\%), consistent with the expectations of the hard state. As the accretion rate increasing, the source entered into the intermediate state. The power-law component became softer and the disk component started to be significant in the flux. Upon closer inspection, there were three parts in this state (Figure~\ref{fig:spe} and \ref{fig:flu}): MJD 58007 -- 58014 (ObsID 10264004 -- 10264010, part 1), MJD 58015 -- 58037 (ObsID 10264011 -- 10264031, part 2) and MJD 58049 -- 58051 (ObsID 88245004 -- 88246001, part3). $\Gamma$ were about $2.5$ in part 2, larger than that of parts 1 and 3 ($\Gamma \sim 2.0$); the inner disk temperature of part 2 had a significantly higher level, while the inner disk radius ($N_{\rm disk}$) were much lower than that of parts 1 and 3; $f_{\rm disk}$ of part 2 were larger, so the disk component was stronger. Although we could not perform the timing analysis with \textit{Swift}/XRT observations due to pileup effect, the timing results of \textit{Insight}-HXMT found that the fractional rms over 0.1--32 Hz (1-12 keV) decreased from $\sim$15\% on MJD 58014 to 1.9\% on MJD 58015 and meanwhile QPOs switched from type C to type B \citep{Huang2018}. Inferred from the spectral and timing properties, part 1 is in hard intermediate (HIM) state, while part 2 is soft intermediate (SIM) state. We also refer to part 3 as a hard intermediate state because the spectral properties are similar to that of part 1, even though there are no timing studies during this period. After a ninety-day interval, when \textit{Swift} observed the source again (MJD 58137), we found that the disk component dominated the emission with $f_{\rm disk} > $ 80\% and $N_{\rm disk}$ was stable, indicating that the source may go through another soft intermediate state, which was not detected due to the lack of observations, and reached the soft state finally. The source stayed in the soft state until MJD 58230. Around MJD 58233--58234 (ObsID 10264046 -- 10264047), $f_{\rm disk}$ decreased to be less than 40\%, and four days later, the source was in the hard state again. We thus revise ObsID 10264046 and 10264047 as intermediate state. For convenience, the state classifications are listed in Table~\ref{tab:mod}.

%%%%%%%%%%%%%%%%%%%%%%%%%%%%%%%%%%%%%%%%%%%%%%%%%%%%%%%%%%%%%%%%%%%%%%%%%%
\begin{figure*}
\centering
\includegraphics[width=0.33\textwidth]{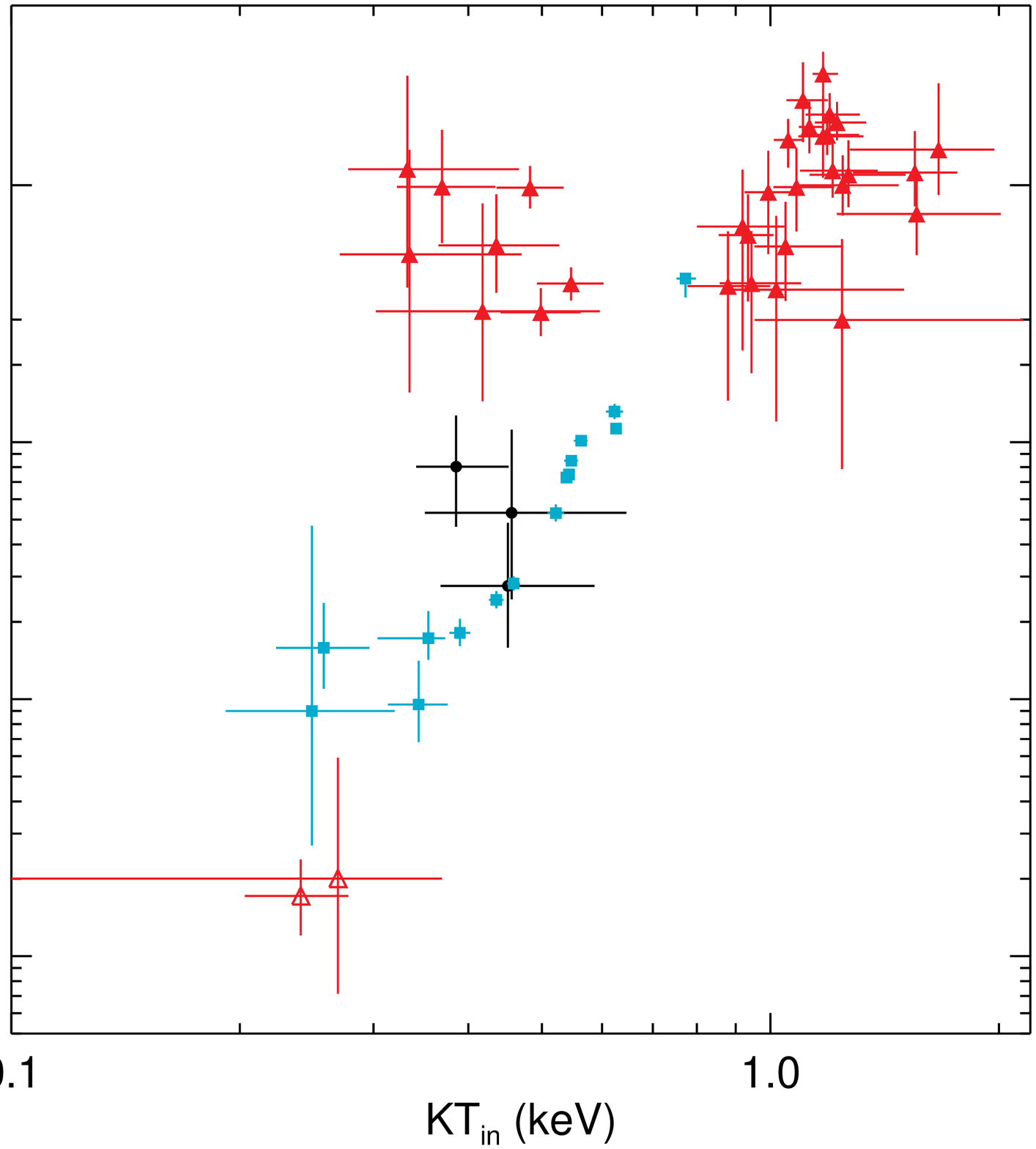}
\includegraphics[width=0.33\textwidth]{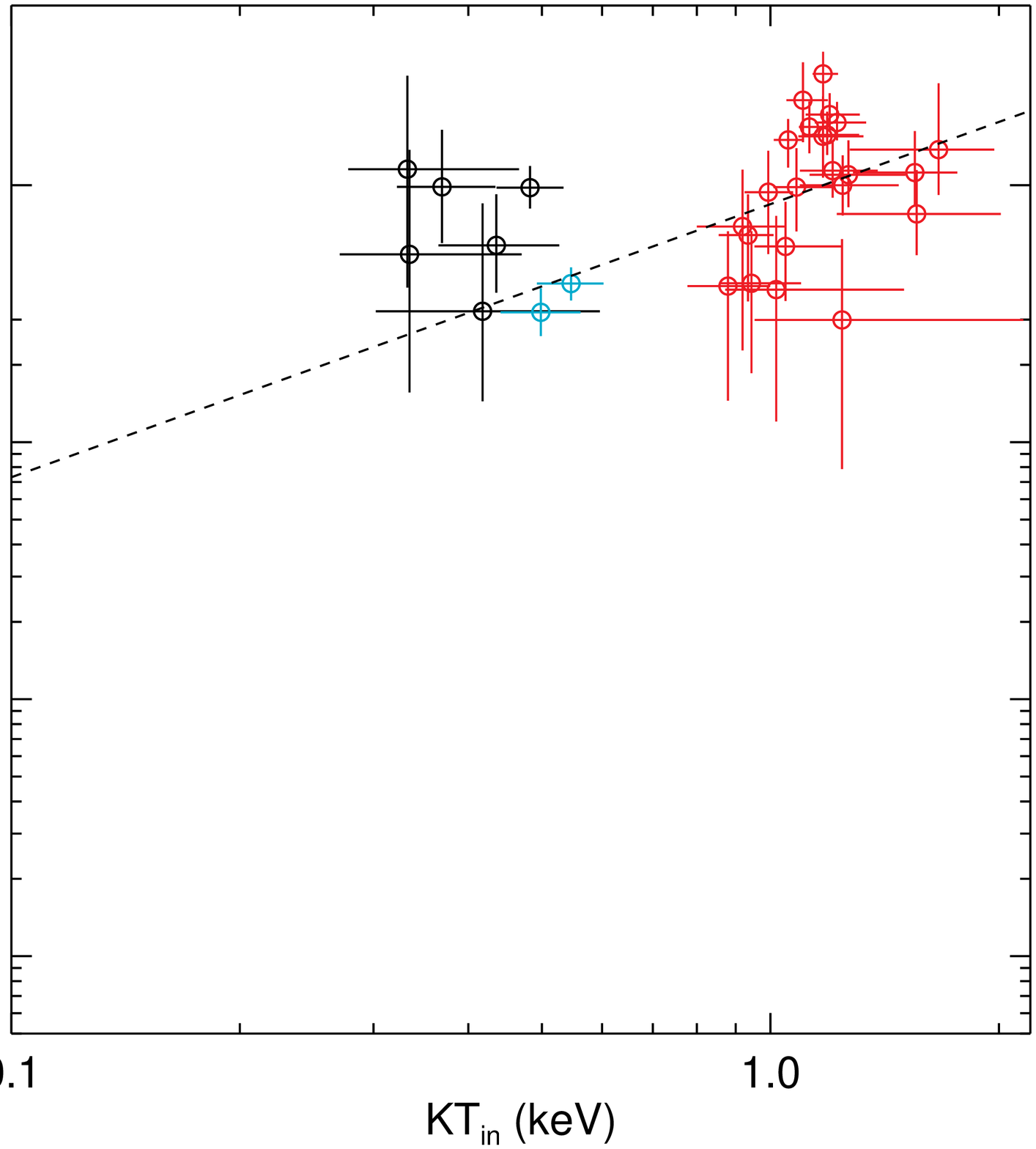}
\includegraphics[width=0.33\textwidth]{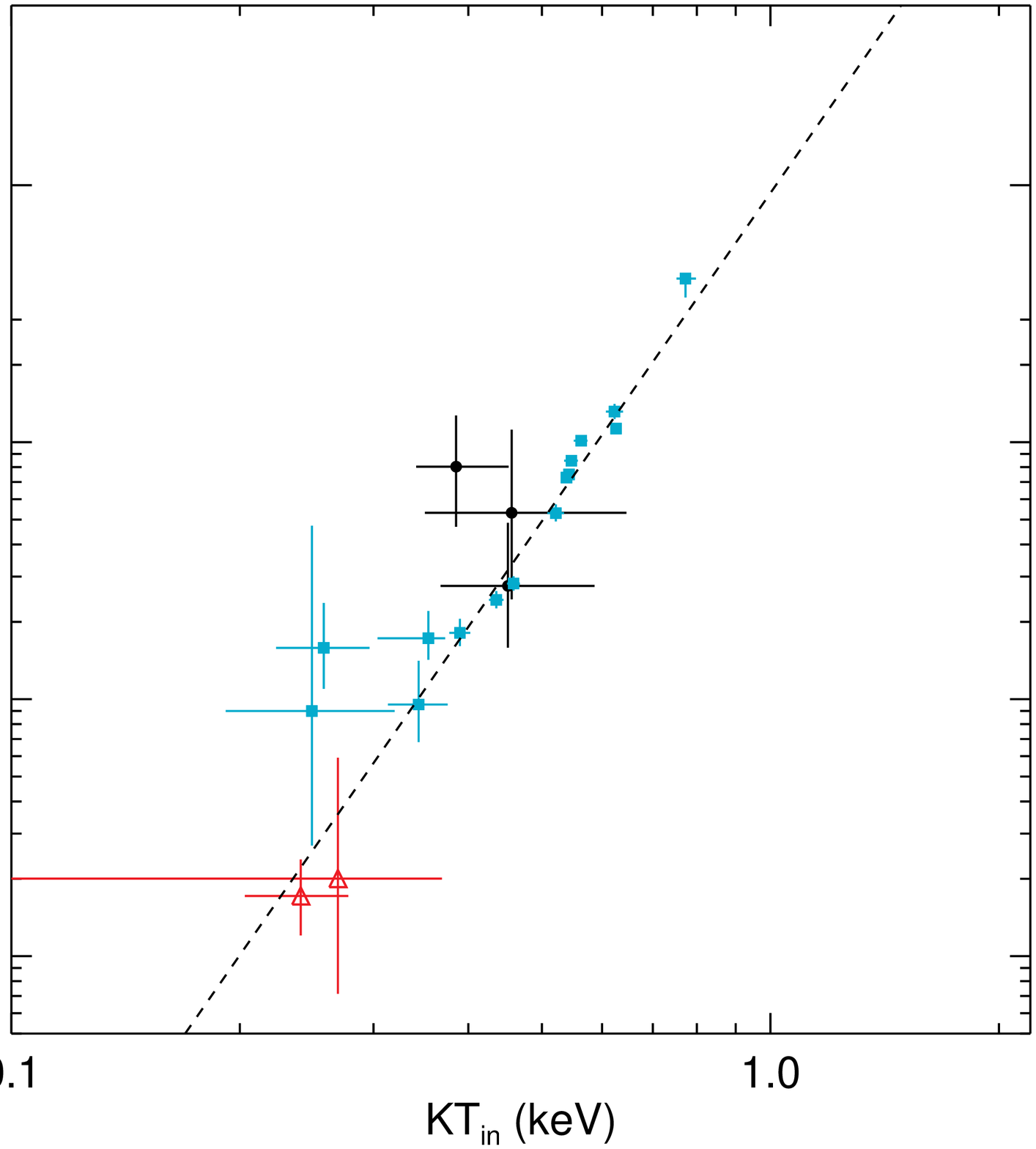}
\caption{The disk luminosity $L_{\rm disk}*D_{10}^{-2}$ vs. the disk inner temperature ($kT_{\rm in }$). The dashed lines represent the best fitting relation. Left panel: the disk component of the whole outburst; different states are marked using the same colored symbols as Figure~\ref{fig:spe}. Middle panel: the disk component in the high intermediate state (red filled triangles in left panel); black -- first HIM state, red -- SIM state, blue -- second HIM state. Right panel: the disk component in the hard (black dots), soft (blue squares) and lower intermediate states (red opening triangles).
\label{fig:ft}}
\end{figure*}
%%%%%%%%%%%%%%%%%%%%%%%%%%%%%%%%%%%%%%%%%%%%%%%%%%%%%%%%%%%%%%%%%%%%%%%%%%

%%%%%%%%%%%%%%%%%%%%%%%%%%%%%%%%%%%%%%%%%%%%%%%%%%%%%%%%%%%%%%%%%%%%%%%%%%
\begin{figure*}
\centering
\includegraphics[width=0.33\textwidth]{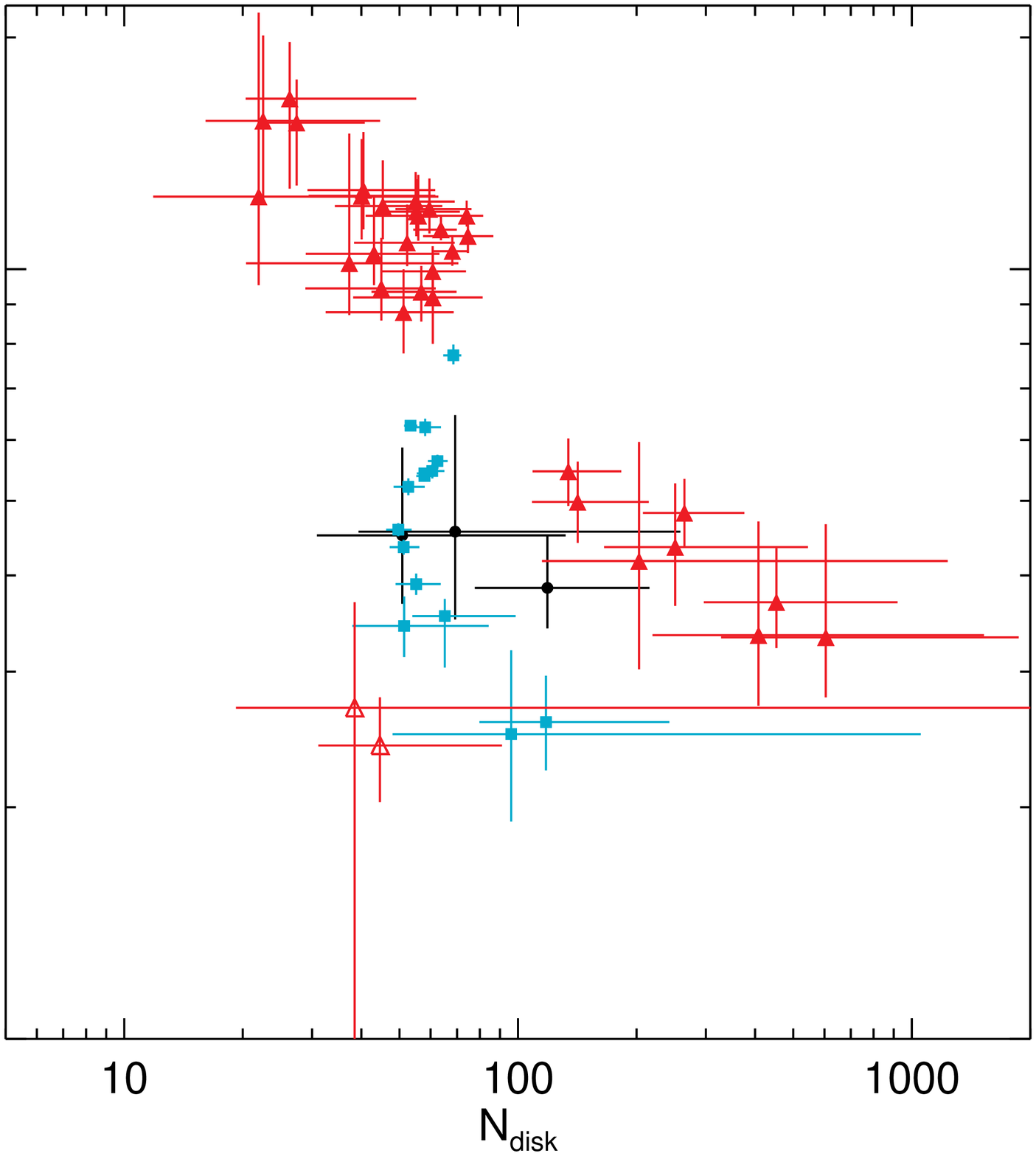}
\includegraphics[width=0.33\textwidth]{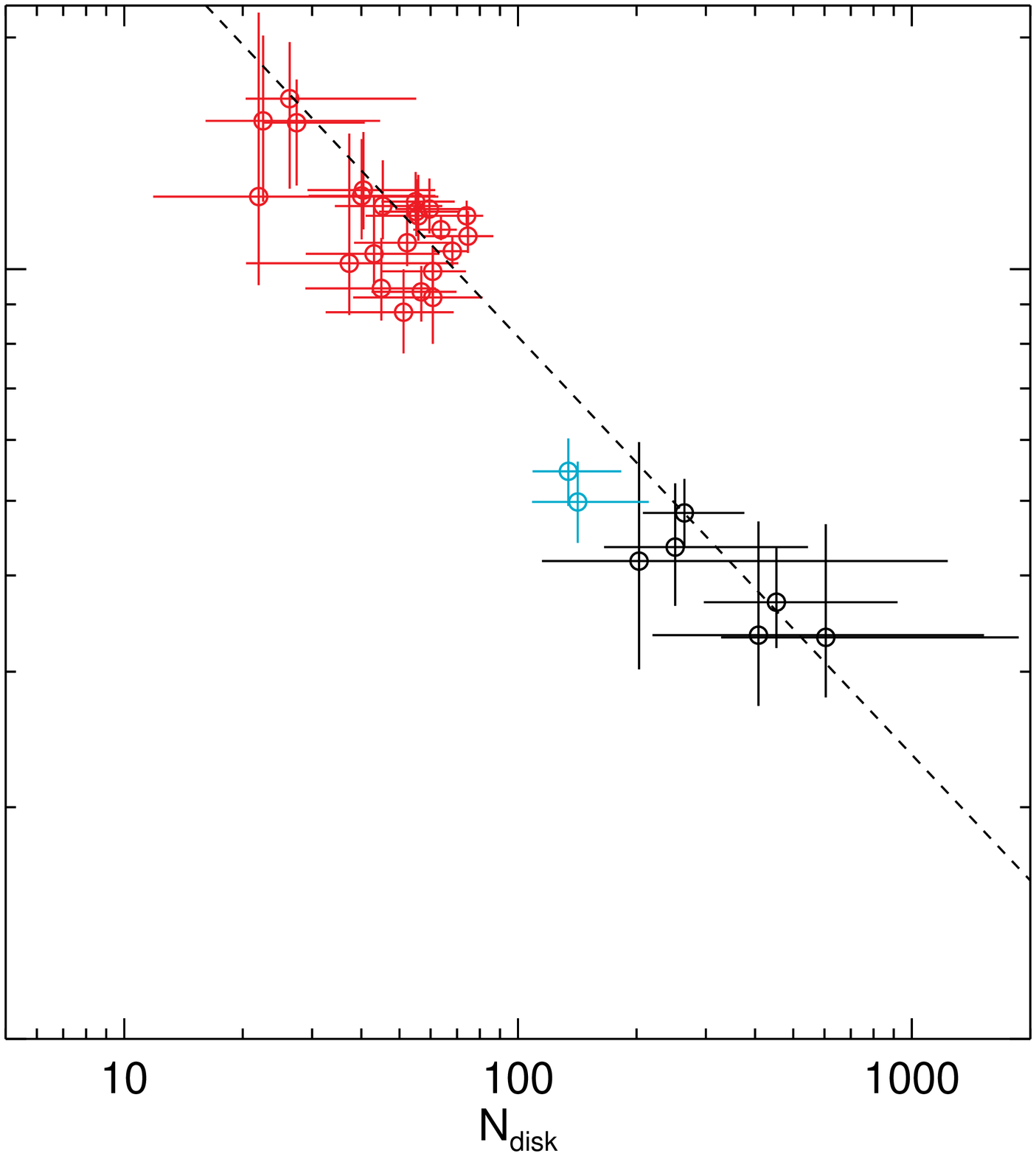}
\includegraphics[width=0.33\textwidth]{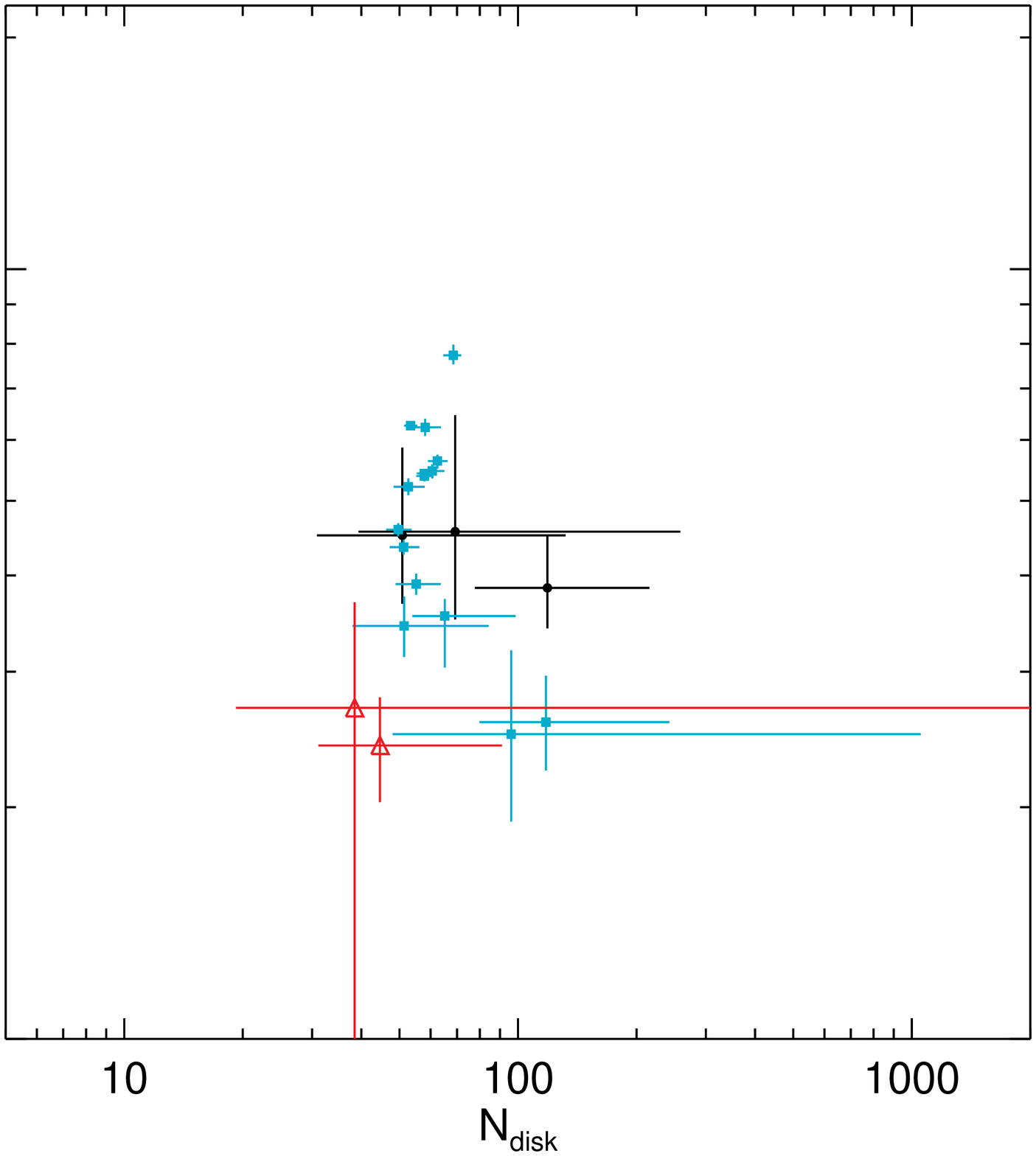}
\caption{Disk inner temperature vs. the square root of the normalization of the diskbb model, $N_{\rm disk} = R_{\rm in}(\rm km)*D_{10}^{-1}*\sqrt{cos(i)}$. The dashed line represents the best fitting relation. Left panel: the disk component of the whole outburst; different states are marked using the same colored symbols as Figure~\ref{fig:spe}. Middle panel: the disk component in the high intermediate state; different states are marked using the same colored symbols as Figure~\ref{fig:ft}. Right panel: the disk component in the hard (black dots), soft (blue squares) and lower intermediate states (red opening triangles).
\label{fig:tr}}
\end{figure*}
%%%%%%%%%%%%%%%%%%%%%%%%%%%%%%%%%%%%%%%%%%%%%%%%%%%%%%%%%%%%%%%%%%%%%%%%%%

\subsection{Disk Component}

In general, the disk component of black hole can be well described with a standard thin disk \citep{Shakura1973}. Its disk luminosity $L_{\rm disk}$ and the inner disk temperature $T_{\rm in }$ follows $L_{\rm disk} \propto T_{\rm in }^4$, and its radial temperature profile varies as $T(R) \propto R^{-0.75}$. However, the disk structure may be modified if the source is bright. The peak flux of MAXI J1535-571 is $\sim 5$~Crab, we thus need to test whether its disk component follows the standard scenario. As some key parameters, i.e, the distance $D$ and the inclination $i$, are not well constrained, we thus use the disk luminosity $L_{\rm disk}*D_{10}^{-2}$ and the square root of the normalization of the diskbb model, $N_{\rm disk} = R_{\rm in}(\rm km)*D_{10}^{-1}*\sqrt{cos(i)}$ instead, to test the relations $L_{\rm disk} \propto T_{\rm in }^{\alpha}$ and $T(R) \propto R^{-p}$.

%L-T relation
$L_{\rm disk}*D_{10}^{-2}$ versus $T_{\rm in}$, as shown in the left panel of Figure~\ref{fig:ft}, follows a tight relation, except for the observations which the source was in the high intermediate state. We thus fit $L_{\rm disk}$ versus $T_{\rm in}$ in the high intermediate state (middle panel of Figure~\ref{fig:ft}) and the remaining states (right panel of Figure~\ref{fig:ft}) to a power-law function, respectively, and found that $\alpha$ is $1.06 \pm 0.14$ and $4.24\pm0.11$. Moreover, we also test the $L_{\rm disk} \propto T_{\rm in }^{\alpha}$ relation if only fitting data in the SIM state and soft state, respectively, and obtain $\alpha=1.0 \pm 0.4$ for the SIM state and $\alpha=4.20 \pm 0.12$ for the soft state.

%R-T relation
$T$ versus $R$ are plotted in Figure~\ref{fig:tr}. In the high intermediate state (middle panel of Figure~\ref{fig:tr}), $N_{\rm disk}$ and $T_{\rm in }$ appear to have a robust correlation. The radial temperature profile has a flatter $p$ of $0.54 \pm 0.03$ if fitting to a power-law function ($T(R) \propto R^{-p}$). The data in the right panel of Figure~\ref{fig:tr} show clearly that the soft state has an approximately constant radius, consistent with the expectation that the inner disk extends to the innermost stable circular orbit (ISCO) in the soft state. $N_{\rm disk}$ in the hard state is not significantly larger than that in the soft state, indicating that the inner disk is not significantly truncated, consistent with the result of \citet{Xu2018}, which report an inner disk radius of $< 2.01~R_{\rm ISCO}$ in the hard state based on the reflection modeling of \textit{NuSTAR} data.

%%%%%%%%%%%%%%%%%%%%%%%%%%%%%%%%%%%%%%%%%%%%%%%%%%%%%%%%%%%%%%%%%%%%%%%%%%
\begin{table*}
\begin{minipage}{160mm}
\centering
\caption{Spectral Fitting of MAXI~J1535-571. PL: power-law model; PLBB: power-law + {\tt diskbb}; BB: {\tt diskbb}. $N_{\rm H}$ is the X-ray absorption column density in units of $10^{22}~\rm atoms~cm^{-2}$; $\Gamma$ is the power-law photon index; $kT_{\rm in}$ is the accretion disk temperature of the {\tt diskbb} model in units of keV; $N_{\rm disk}$ is the square root of the normalization of the diskbb model, $N_{\rm disk} = R_{\rm in}(\rm km)*D_{10}^{-1}*\sqrt{cos(i)}$; All errors and limits are at 90\% confidence level. }
\label{tab:mod}
\begin{tabular}{lllllll}
\hline
ObsID  & Model & $N_{\rm H}$ & $\rm \Gamma$ & $kT_{in}$  & $N_{\rm disk}$ & $\chi^2/{\rm dof}$ \\
 &  & ($10^{22}~\rm cm^{-2}$) &  & (keV) &  & ($C^2/{\rm dof}$)$^f$ \\
\hline
770431000 & PL & $3.6\pm0.2$ & $1.50\pm0.07$ & \nodata & \nodata  & 410.9/450 \\
770502000 & PL & $3.5^{+0.7}_{-0.6}$ &  $1.4\pm0.2$ & \nodata & \nodata   & 49.1/58 \\
770656000 & PL & $3.2^{+0.8}_{-0.7}$ &  $1.4\pm0.3$ & \nodata & \nodata   & 45.0/42 \\
770656001 & PLBB  & $4.0\pm0.3$ & $1.34^{+0.06}_{-0.07}$ & $0.45^{+0.14}_{-0.08}$ & $50^{+80}_{-20}$ &   736.0/700 \\
10264002 & PLBB  & $3.9^{+0.5}_{-0.4}$& $1.34^{+0.09}_{-0.11}$  & $0.46^{+0.19}_{-0.11}$ & $70^{+190}_{-30}$ &   583.6/608  \\
10264003 & PLBB  & $3.8\pm0.2$& $1.42\pm0.02$  & $0.39^{+0.07}_{-0.04}$ & $120^{+100}_{-40}$  & 984.4/855  \\
771371000 & PL  & $3.3\pm0.3$& $1.32\pm0.10$  & \nodata & \nodata &   188.6/227 \\
10264004 &  PL & $3.76\pm0.06$& $1.97\pm0.03$  & \nodata &\nodata  &   763.0/696 \\
10264005 & PLBB  & $4.4^{+0.3}_{-0.2}$ & $1.83^{+0.07}_{-0.08}$ & $0.44^{+0.09}_{-0.07}$ & $250^{+290}_{-90}$ &   695.6/672 \\
10264007 & PL  & $3.99\pm0.08$& $2.13\pm0.03$  & \nodata & \nodata &    701.5/661 \\
10264006 & PLBB  & $4.2^{+0.4}_{-0.3}$ & $2.03\pm0.06$ & $0.33^{+0.14}_{-0.06}$ & $410^{+1120}_{-190}$ &   700.9/681  \\
10264008 & PLBB  & $4.2^{+0.4}_{-0.2}$  & $2.00^{+0.07}_{-0.08}$ & $0.42^{+0.18}_{-0.12}$ & $200^{+1030}_{-90}$ &  681.5/682 \\
10264009 & PLBB  & $4.7\pm0.3$  &$1.98\pm0.06$ & $0.37^{+0.06}_{-0.05}$ & $450^{+470}_{-160}$  & 781.1/689 \\
88245001 & PLBB  & $4.45^{+0.15}_{-0.13}$  &$1.84\pm0.05$ & $0.48^{+0.05}$ & $260^{+110}_{-60}$ &   859.2/769 \\
10264010 & PLBB  & $4.8^{+0.4}_{-0.3}$ & $2.17^{+0.06}_{-0.07}$  & $0.33^{+0.13}_{-0.05}$ & $600^{+1300}_{-300}$ &   718.1/692 \\
10264011 & PLBB  & $4.9^{+0.6}_{-0.5}$& $2.6^{+0.9}_{-0.4}$  & $1.7^{+0.3}_{-0.4}$ & $26^{+29}_{-6}$ &  486.1/516 \\
88245002 & PLBB  &$5.0^{+0.2}_{-0.3}$ &  $2.34^{+0.15}_{-0.16}$ & $1.2^{+0.9}_{-0.3}$ & $22^{+41}_{-10}$ &   687.8/660 \\
88245003 & PLBB  &$5.1\pm0.3$ & $2.5^{+0.3}_{-0.2}$  & $1.6^{+0.5}_{-0.3}$ & $23^{+22}_{-6}$   & 709.8/645 \\
10264012 & PLBB & $4.7\pm0.3$& $2.4^{+0.2}_{-0.3}$  & $1.22^{+0.11}_{-0.08}$ & $55^{+14}_{-10}$ &    789.9/648\\
10264013 & PLBB  & $3.9^{+0.4}_{-0.2}$& $1.8^{+0.5}_{-1.9}$  & $1.17^{+0.05}_{-0.04}$ & $74^{+7}_{-11}$   & 623.4/615 \\
10264014 & PLBB  & $4.3^{+0.5}_{-0.3}$& $1.9^{+0.5}_{-1.6}$  & $1.10^{+0.09}_{-0.05}$ & $75^{+12}_{-17}$   & 597.5/577 \\
10264015 & PLBB  & $5.7^{+0.4}_{-0.3}$& $3.0^{+0.5}_{-0.3}$  & $1.5^{+0.2}_{-0.3}$ & $27^{+13}_{-5}$ &  678.8/618 \\
10264016 & PLBB  & $4.9^{+0.3}_{-0.4}$& $2.3^{+0.2}_{-0.3}$  & $1.27^{+0.24}_{-0.14}$ & $40^{+21}_{-11}$ &   658.2/634 \\
10264017 & PLBB  & $5.1^{+0.3}_{-0.4}$& $2.6^{+0.3}_{-0.4}$  & $1.20^{+0.11}_{-0.08}$ & $60^{+17}_{-11}$ &   650.8/611 \\
10264018 & PLBB  & $5.6\pm0.3$& $2.7\pm0.2$  & $1.25^{+0.23}_{-0.15}$ & $40^{+22}_{-11}$ &   636.2/632 \\
10264019 & PLBB  & $4.8^{+0.3}_{-0.4}$& $2.5^{+0.2}_{-0.3}$  & $1.19^{+0.12}_{-0.08}$ & $55^{+16}_{-11}$ &  637.5/622 \\
10264020 & PLBB  &$4.2\pm0.3$ & $2.0^{+0.2}_{-0.5}$  & $1.17^{+0.15}_{-0.08}$ & $56^{+19}_{-15}$ &   658.5/634 \\
10264021 & PLBB  &$4.0\pm0.3$ & $1.9^{+0.3}_{-0.4}$  & $0.99^{+0.08}_{-0.07}$ & $61^{+13}_{-16}$ &   644.7/632 \\
10264022 & PLBB  &$3.7\pm0.2$ & $1.6^{+0.3}_{-0.5}$  & $1.06^{+0.05}_{-0.04}$ & $68^{+6}_{-8}$ &   699.5/651 \\
10264023 & PLBB  &$4.4^{+0.3}_{-0.4}$ & $2.2^{+0.2}_{-0.4}$  & $1.08^{+0.13}_{-0.07}$ & $52^{+17}_{-14}$ &   607.4/631 \\
10264024 & PLBB  &$3.8\pm0.2$ & $1.9^{+0.2}_{-0.3}$  & $0.93^{+0.08}$ & $57^{+13}_{-14}$ &   678.7/645 \\
10264025 & PLBB  &$4.8\pm0.4$ & $2.1^{+0.3}_{-0.4}$  & $0.92^{+0.13}_{-0.12}$ & $60\pm20$ &   612.3/612 \\
10264026 & PLBB &$3.9^{+0.3}_{-0.2}$  & $1.8^{+0.5}_{-1.2}$ & $1.13^{+0.04}$ &$64^{+6}_{-9}$    & 658.8/645 \\
10264027 & PLBB  &$4.7^{+0.2}_{-0.3}$ & $2.3\pm0.2$  & $1.02^{+0.48}_{-0.15}$ & $37^{+33}_{-17}$ &   681.4/656 \\
10264028 & PLBB  &$4.8^{+0.3}_{-0.4}$  & $2.5^{+0.2}_{-0.3}$ & $1.21^{+0.18}_{-0.11}$ & $45^{+19}_{-11}$ &   599.2/618 \\
10264029 & PLBB  &$4.1\pm0.3$  & $2.2^{+0.2}_{-0.3}$ & $0.94^{+0.15}_{-0.09}$ & $45^{+17}_{-16}$ &   689.1/654 \\
10264030 & PLBB  &$4.6\pm0.3$  & $2.4\pm0.2$ & $1.05^{+0.20}_{-0.09}$ & $43^{+20}_{-14}$ &   656.0/644 \\
10264031 & PLBB  &$4.0\pm0.3$  & $2.1^{+0.2}_{-0.3}$ & $0.88^{+0.12}_{-0.10}$ & $51^{+17}_{-19}$ &   581.9/621 \\
88245004 & PLBB  &$4.72^{+0.16}_{-0.15}$  & $1.93^{+0.07}_{-0.08}$ & $0.55^{+0.06}_{-0.05}$ & $130^{+40}_{-30}$ &   885.3/760 \\
88246001 & PLBB  &$4.23^{+0.17}_{-0.15}$  & $1.90^{+0.06}_{-0.07}$ & $0.50\pm0.06$ & $140^{+60}_{-40}$ &  694.2/729 \\
10491002 & PLBB  &$4.21^{+0.55}_{-0.11}$  & $2.0^{+1.5}_{-1.8}$ & $0.77\pm0.02$ & $69^{+3}_{-4}$ &  508.7/510 \\
10264032 &  BB  &  $3.45 \pm 0.15$ & \nodata  & $0.623 \pm 0.016$ & $58^{+6}_{-5}$ &  241.1/253 \\
10264033 &  PLBB  & $3.27 \pm 0.06$  & $1.0^{+4.6}_{-0.0}$$^c$  & $0.626^{+0.008}_{-0.007}$  & $53 \pm 2$  & 448.2/404  \\
10264034 &  PLBB  & $3.41^{+0.16}_{-0.08}$  & $1.5^{+2.2}_{-2.5}$  & $0.563^{+0.011}_{-0.013}$  &  $62^{+4}_{-3}$ & 381.3/371  \\
10264035 &  PLBB  & $3.42 \pm 0.10$  & $1.0^{+2.3}_{-0.0}$$^c$  & $0.547^{+0.011}_{-0.012}$  & $61 \pm 4$  & 364.9/323  \\
10264036 &  PLBB  & $3.26^{+0.06}_{-0.07}$  & $1.0^{+1.5}_{-0.0}$$^c$  & $0.543 \pm 0.007$  &  $58^{+3}_{-2}$ & 391.5/353  \\
10264037 &  PLBB  & $3.31 \pm 0.07$  & $1.0^{+1.5}_{-0.0}$$^c$  &  $0.539^{+0.008}_{-0.009}$ & $58 \pm 3$  & 442.4/347  \\
10264038 &  BB  & $3.12^{+0.14}_{-0.13}$  & \nodata  & $0.522 \pm 0.013$  &  $53^{+5}_{-4}$ & 245.3/226  \\
10264039 &  BB  & $2.87 \pm 0.10$  & \nodata  & $0.459 \pm 0.009$  & $50^{+4}_{-3}$  &  275.5/247 \\
10264040 &  BB  & $2.95^{+0.12}_{-0.11}$  & \nodata  & $0.435 \pm 0.010$  &  $51^{+5}_{-4}$ & 209.2/227  \\
10264041 &  BB  & $2.88^{+0.17}_{-0.16}$  &  \nodata & $0.390 \pm 0.012$  &  $55^{+8}_{-6}$ & 184.0/171  \\
10264042 &  PLBB  & $2.96^{+0.25}_{-0.23}$  & $1.0^{+5.6}_{-0.0}${$^c$}  & $0.354^{+0.019}_{-0.05}$  &  $65^{+33}_{-11}$ & 136.6/133  \\
\hline
\end{tabular}
\end{minipage}
\end{table*}

%%%%%%%%%%%%%%%%%%%%%%%%%%%%%%%%%%%%%%%%%%%%%%%%%%%%%%%%%%%%%%%%%%%%%%%%%%

%%%%%%%%%%%%%%%%%%%%%%%%%%%%%%%%%%%%%%%%%%%%%%%%%%%%%%%%%%%%%%%%%%%%%%%%%%
\begin{table*}
\begin{minipage}{160mm}
\centering
\contcaption{}
%\label{tab:mod}
\begin{tabular}{lllllll}
\hline
ObsID  & Model & $N_{\rm H}$ & $\rm \Gamma$ & $kT_{in}$  & $N_{\rm disk}$ & $\chi^2/{\rm dof}$ \\
 &  & ($10^{22}~\rm cm^{-2}$) &  & (keV) &  & ($C^2/{\rm dof}$)$^f$ \\
\hline
10264043 &  BB  & $3.04^{+0.45}_{-0.40}$  & \nodata  & $0.34 \pm 0.03$  & $51^{+33}_{-13}$  & 100.9/63  \\
10264044 &  BB  & $3^b$  & \nodata  & $0.26^{+0.04}_{-0.03}$  & $120^{+120}_{-40}$  & 21.5/11  \\
10264045 &  BB  & $3.2^{+1.5}_{-1.1}$  & \nodata  & $0.25^{+0.07}_{-0.06}$  & $100^{+960}_{-50}$  & 22.1/21  \\
10264046 &  PLBB  & $3.0^{+4.8}_{-0.9}$  & 1$^c$$^e$  & $0.27^{+0.10}_{-0.27}$  & $40^{+2030}_{-20}$  &  11.1/9 \\
10264047 &  PLBB  & $3^b$  & $1.0^{+8.7}_{-0.0}$$^c$  & $0.24 \pm 0.04$  & $45^{+46}_{-13}$  & 6.6/5  \\
10264048$^a$ &  PL  & $2.4^{+1.7}_{-1.3}$  & $2.6^{+1.1}_{-1.0}$  & \nodata  & \nodata  & 58.7/63  \\
10264049$^a$ &  PL  & $2.8^{+2.8}_{-1.8}$  & $1.1^{+0.9}_{-0.8}$  & \nodata  & \nodata  & 55.3/68  \\
10264050$^a$ &  PL  & $3^b$  & $0.8^{+4.5}_{-0.0}$$^d$  &  \nodata & \nodata  & 0.3/1  \\
10264051$^a$ &  PL  & $3^b$  & $0.3^{+1.5}_{-1.9}$  & \nodata  & \nodata  & 4.3/10  \\
10264052$^a$ &  PL  & $3^b$  & $0.4^{+2.2}_{-0.0}$$^d$  & \nodata  & \nodata  & 3.2/4  \\
10264053$^a$ &  PL  & $4.2^{+2.0}_{-1.6}$  & $1.7 \pm 0.6 $  & \nodata  & \nodata  & 120.6/159  \\
\hline
\end{tabular}
\footnotesize{
\begin{flushleft}
$^a$Using the C statistic (cstat in xspec) to fit the spectra. \\
$^b${$N_{\rm H}$} is fixed at $3 \times 10^{22}~\rm cm^{-2}$ as it is difficult to be constrained. \\
$^c$The power-law component is weak and $\Gamma$ favors a negative value when it is a free fit parameter. So, the lower limit of $\Gamma$ is fixed to 1. \\
$^d$The parameter lower limit and$^e$upper limit are invalid due to a low statistic. \\
$^f$The goodness of fit in C statistic is described in terms of $C^2$.
\end{flushleft} }
\end{minipage}
\end{table*}

%%%%%%%%%%%%%%%%%%%%%%%%%%%%%%%%%%%%%%%%%%%%%%%%%%%%%%%%%%%%%%%%%%%%%%%%%%

\section{Discussion}
\label{sec:dis}

\textit{Swift} monitoring observations and the detailed spectral analyses of the recently discovered BH candidate, MAXI~J1535-571, are presented in this paper. 

\subsection{State evolution}
The source was first detected in the hard state, then went through the high intermediate state, and reached the soft state afterward. Near the end of the outburst, the source went back to the low intermediate state and finally returned to the hard state\footnote{Some most recent data reveal that the source has several state transitions at the end of the outburst \citep{ATel11652, ATel11682}. But the source flux is weak, we thus decide not to do detailed spectral analysis. }. It shows a specific counter clockwise Q-shaped loop in the HIDs.

We track the evolution of the model parameters, such as $N_{\rm H}$, $\Gamma$, $T_{\rm in}$, $N_{\rm disk}$ and the component flux, during the outburst. $N_{\rm H}$ is larger than the Galactic value, suggesting that there are some intrinsic materials to obscure the source. Especially, $N_{\rm H}$ shows a peak around the peak of the outburst. We discuss the indication in Section 4.3. The power-law component is weak in the soft state and is significant and harder in the hard state, while the disk component dominates the X-ray spectrum in the soft state and is weaker in the intermediate and hard states, consistent with the typical behaviors of Galactic black hole binaries.  

\subsection{Mass estimation}
The spin of BH can be measured by fitting the disk continuum \citep{Zhang1997}, assuming a standard thin disk, or by fitting the general relativistic ray \citep[e.g.,][]{Garca2014}, and the normalization of the {\tt diskbb} model can be used to infer the BH mass if giving the spin and inclination values. In the soft state, $N_{\rm disk}$ of MAXI~J1535-571 is approximately constant, indicating that the inner disk extends to the ISCO. We thus use the measurements obtained in the soft state to estimate the BH mass. Assuming $a>0.84$, $i=57^{\circ}$ and $R_{\rm in}(\rm km)*D_{10}^{-1}*\sqrt{cos(i)}=50$, the BH mass is obtained to be larger than $14~M_\odot$ for a distance of 10~kpc, or greater than $7~M_\odot$ for a distance of 5~kpc. But we also note that the measurements of $i$ and $a$ are strongly dependent on the model, and the distance is unknown, thus the mass estimation here is only an estimate.

\subsection{Eddington accretion}
\label{sec:edd}

\begin{figure}
\centering
\includegraphics[width=0.49\textwidth]{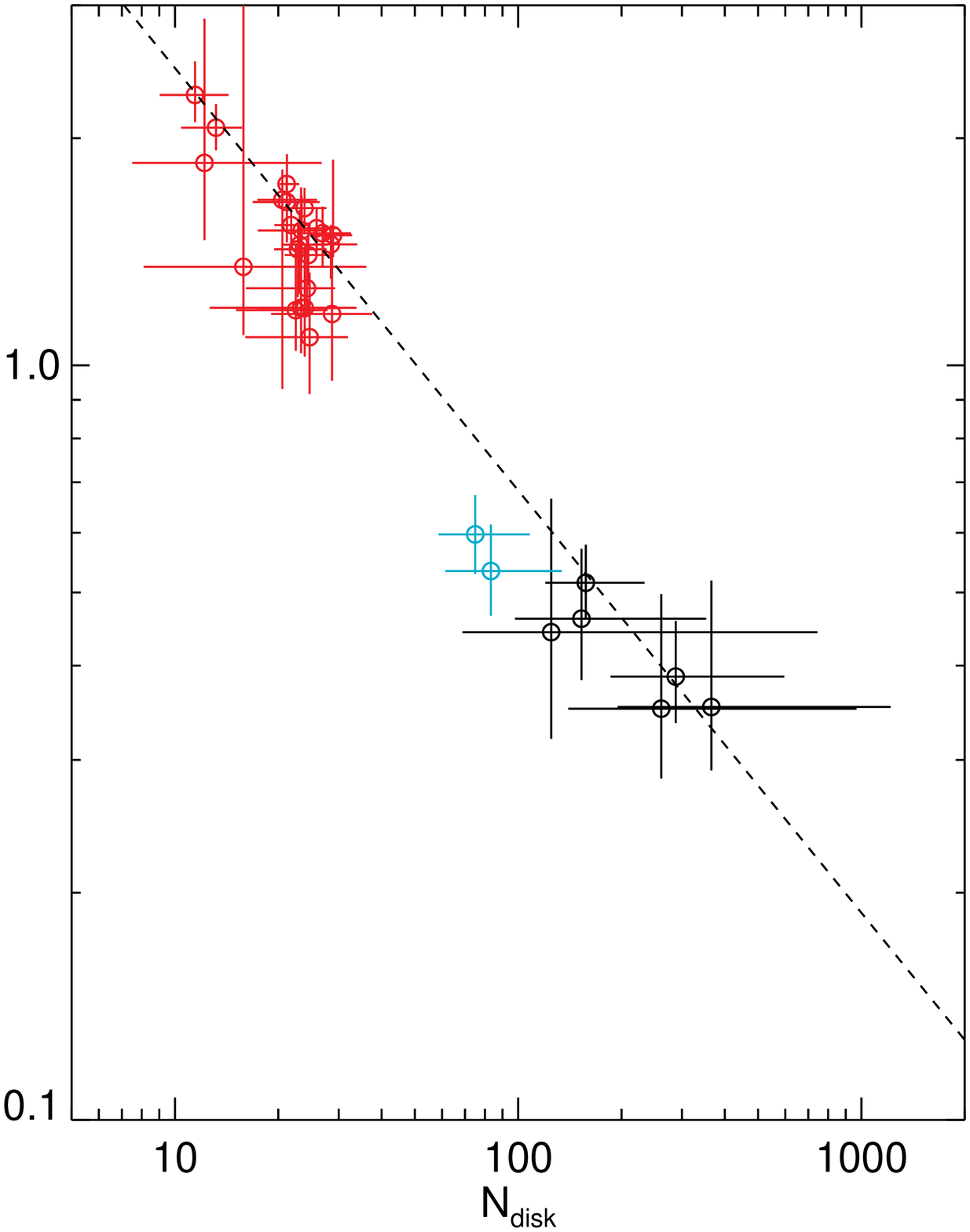}
\caption{Disk inner temperature vs. the square root of the normalization of the {\tt diskpbb} model. Different intermediate states are marked using the same colored symbols as Figure~\ref{fig:ft}.
\label{fig:pbb}}
\end{figure}

The peak flux of the source is up to $\sim 5$~Crab and its disk structure may be modified from the standard thin disk due to the high accretion rate. In order to investigate the physics of the accretion disk, we have examined the relationship between disk luminosity and temperature, $L_{\rm disk} \propto T_{\rm in }^{\alpha}$, and the radial emission profile of the disk temperature, $T(R) \propto R^{-p}$. We found $\alpha \sim 1$ and $p \sim 0.5$ in the high intermediate state (middle panels of Figures~\ref{fig:ft} and \ref{fig:tr}), in contrast with the expectations of a simple thin disk, but consistent with a slim disk \citep{Abramowicz1988}. In the remaining states, an $\alpha$ of $4.24\pm0.11$ well agrees with the value of a standard thin disk. Therefore, we replace the {\tt diskbb} model in the high intermediate state with a disk model with a variable exponent to the disk temperature profile, i.e., $p$-free disk model ({\tt diskpbb})\footnote{https://heasarc.gsfc.nasa.gov/xanadu/xspec/manual/node164.html} in XSPEC. The $p$ expected for a standard thin disk is 0.75, while for a slim disk, is 0.5. If a power law is added to the $p$-free disk model, we are not able to constrain $p$ in the spectral fits, and all possible $p$ values, from 0.5 to 1.0, are allowed. If the $p$-free model is used alone, $p$ appears to be $\sim 0.5$, consistent with the expectations of a slim disk. If we fix $p$ of {\tt diskpbb} at 0.5 and fit the spectra again\footnote{It is difficult to constrain the power-law index for some observations, and a negative value is obtained if fitting the component freely. We thus set the lower limit of the power-law index to be 1.}, $T_{\rm in }$ and $N_{\rm disk}$ would be subject to a power-law function with an index of $0.63 \pm 0.03$ (Figure~\ref{fig:pbb}). If only fitting the data in the SIM state, the index would be $0.56 \pm 0.09$. Moreover, we note that $N_{\rm disk}$ in the SIM state is smaller than that of the soft state, which is seemingly unreasonable because the inner disk already extends to the ISCO in the soft state. However, for a high accretion rate, the advection-dominated flow does not abruptly decrease inside the ISCO. An amount of matter inside the ISCO can produce a substantial radiation, thus the derived inner disk radius is smaller than that of the ISCO \citep{Watarai2000}.

%%%%%%%%%%%%%%%%%%%%%%%%%%%%%%%%%%%%%%%%%%%%%%%%%%%%%%%%%%%%%%%%%%%%%%%%%%
%%%%%%%%%%%%%%%%%%%%%%%%%%%%%%%%%%%%%%%%%%%%%%%%%%%%%%%%%%%%%%%%%%%%%%%%%%

The slim disk is dominated by the advection within the disk, and becomes radiatively inefficient due to the advection process, resulting in the flattening of the $L_{\rm disk}$ and $T_{\rm in}$ relationship. $p$, in the radial emission profile, is expected as $0.5$ for a slim disk \citep{Watarai2000}. A slim disk exists when the luminosity is close to the Eddington or moderately super-Eddington. The disk luminosity of MAXI~J1535-571, in the high intermediate state, is larger than $10^{38}$~\ergs\ and could reach up to several of $10^{39}$~\ergs, approaching to or exceeding the Eddington limit of a stellar mass BH. As a result of this high accretion rate, the disk of MAXI~J1535-571 cannot maintain its thin geometry, and becomes slim. The slim accretion disk has also been suggested for a few Galactic BH binaries in high state, such as XTE J1550-564 \citep{Kubota2004}, 4U 1630-47 \citep{Tomsick2005} and V404 Cyg \citep{Motta2017}, and also for some Ultraluminous X-ray sources \citep{Kaaret2017}.

%%%%%%%%%%%%%%%%%%%%%%%%%%%%%%%%%%%%%%%%%%%%%%%%%%%%%%%%%%%%%%%%%%%%%%%%%%
\begin{figure}
\centering
\includegraphics[width=0.49\textwidth]{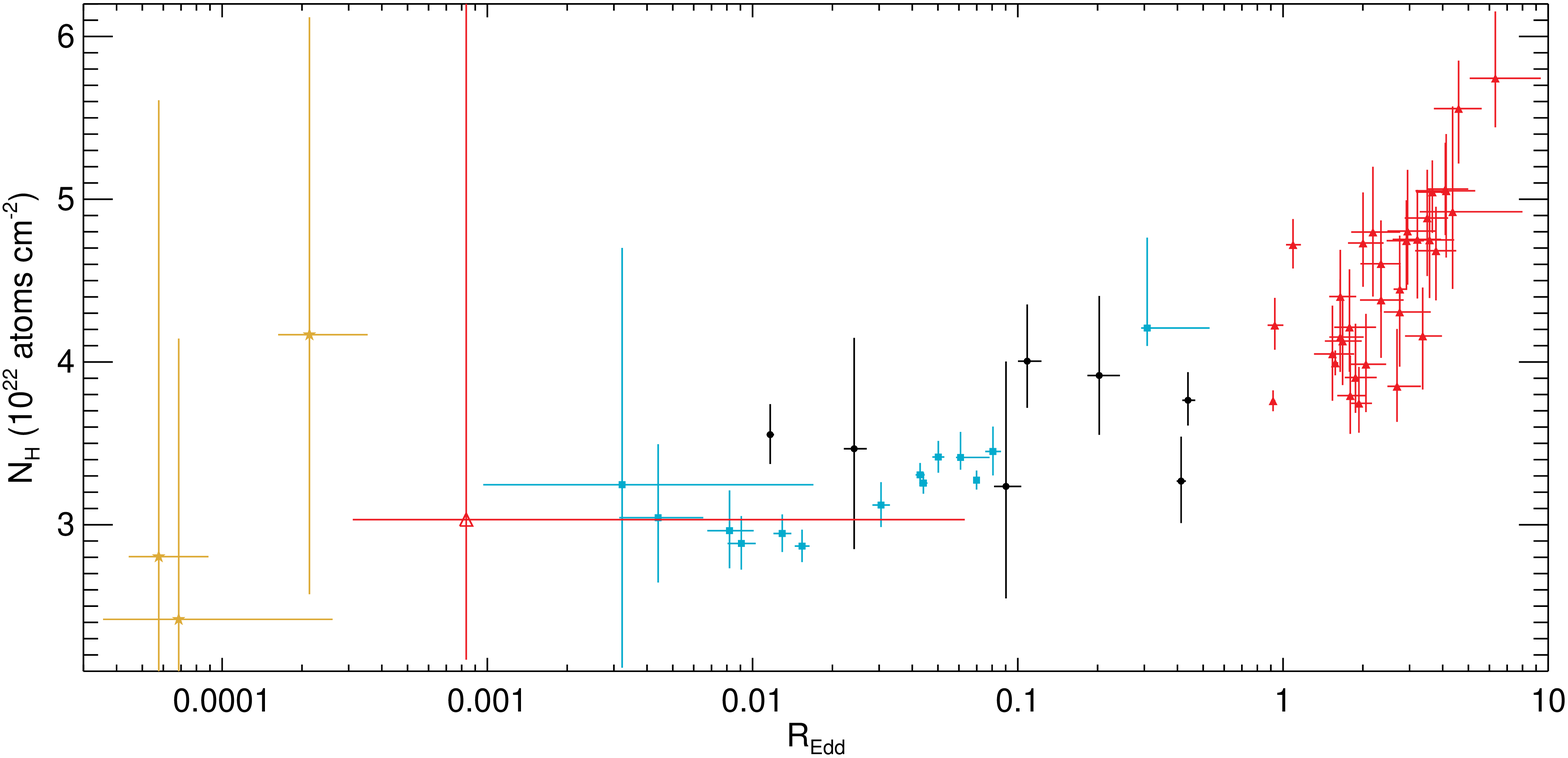}\\
\includegraphics[width=0.49\textwidth]{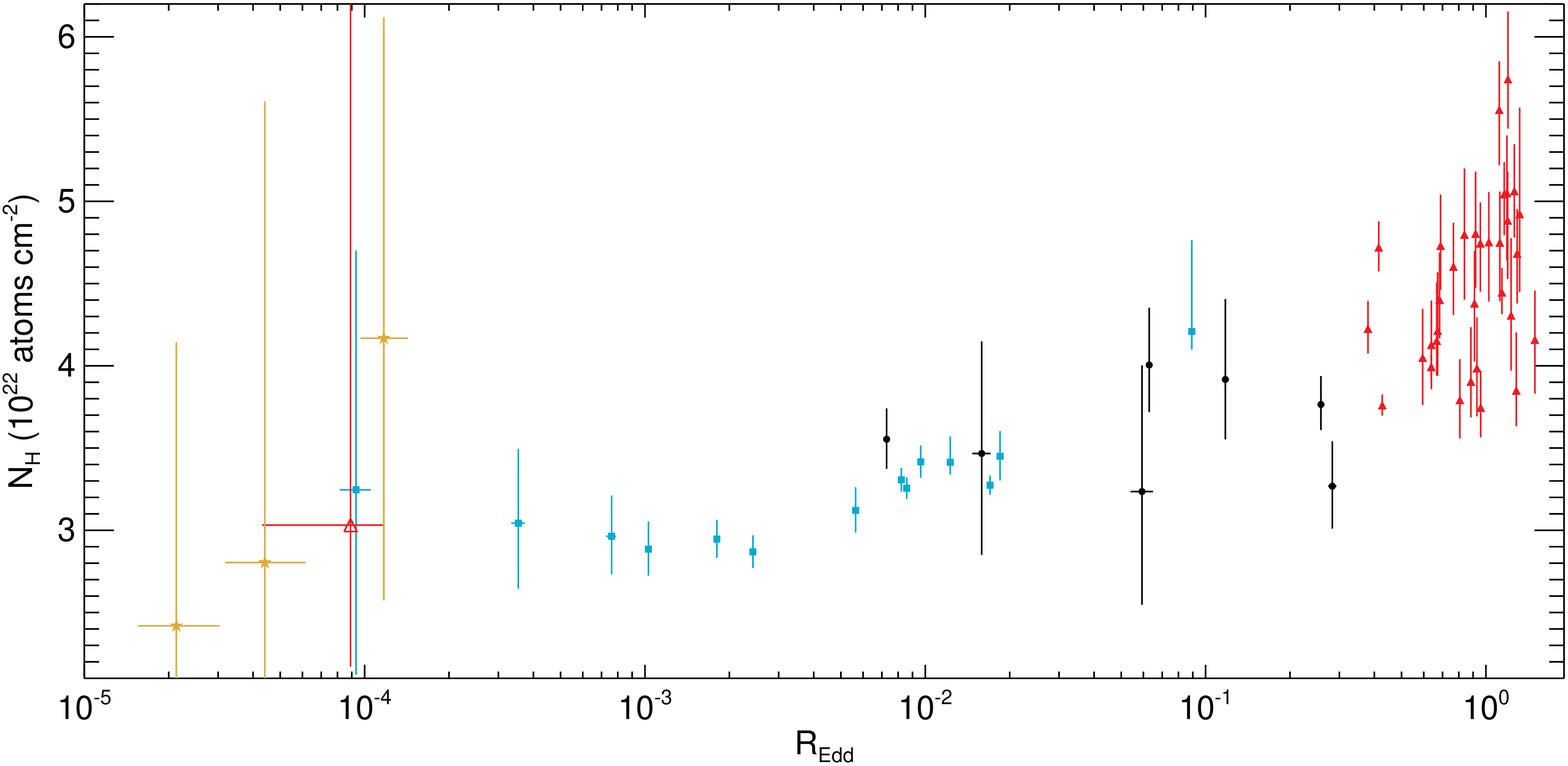}\\
\includegraphics[width=0.49\textwidth]{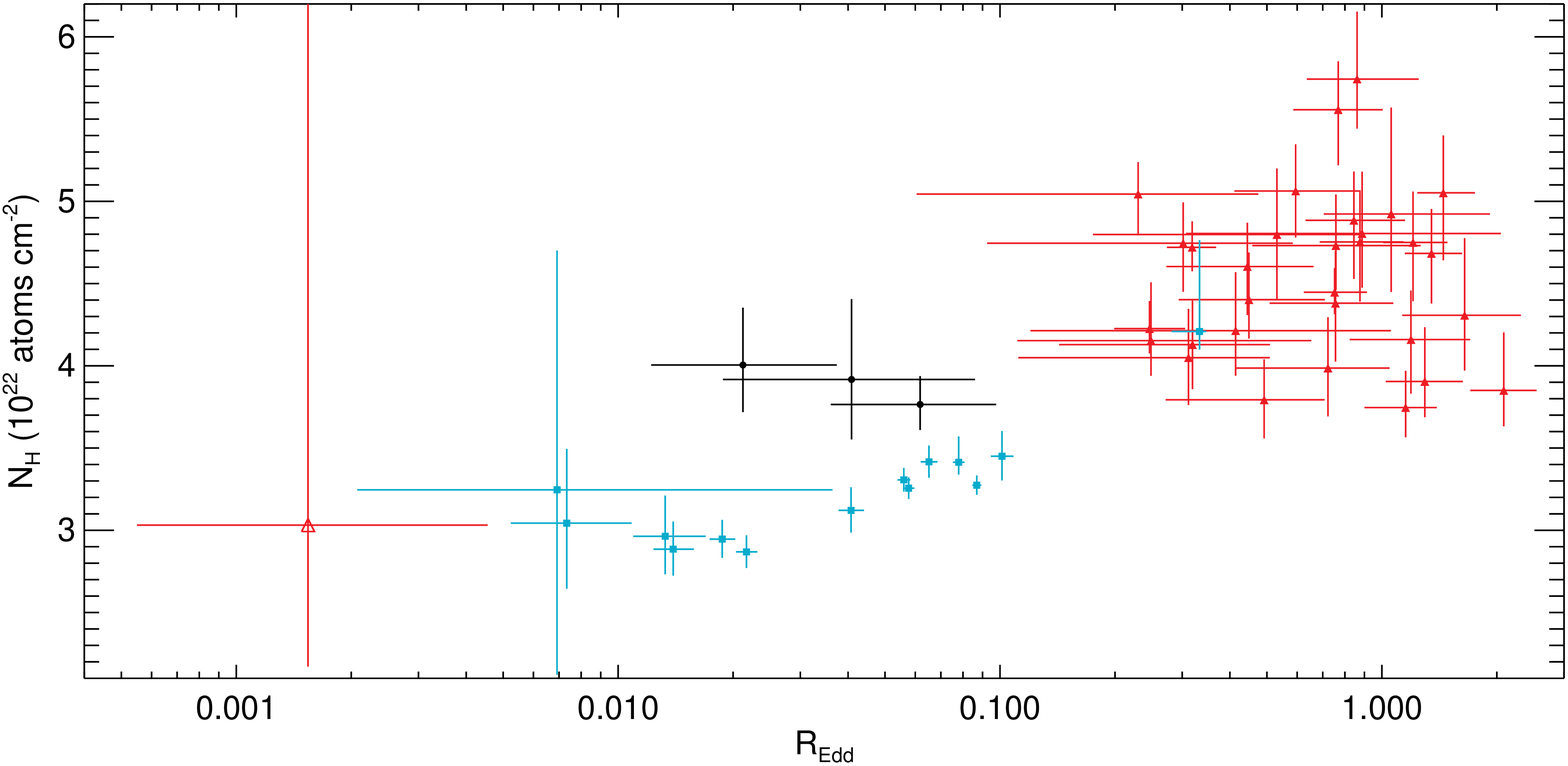}\\
\caption{$N_{\rm H}$ vs. Eddington ratio. upper: unabsorbed source Eddington ratio in the 0.5--10.0 keV band; middle: observed source Eddington ratio in the 0.5--10.0 keV band; lower: unabsorbed disk Eddington ratio. Different states are marked using the same colored symbols as Figure~\ref{fig:spe}.
\label{fig:nh}}
\end{figure}
%%%%%%%%%%%%%%%%%%%%%%%%%%%%%%%%%%%%%%%%%%%%%%%%%%%%%%%%%%%%%%%%%%%%%%%%%%

%nH
In usually, as the mass accretion rate increases, the radiation pressure begins to dominate and a outflow is blown from the disk, which can obscure the source and increase $N_{\rm H}$. Actually, $N_{\rm H}$ in MAXI J1535-571 is larger than the Galactic column density, and shows an evolution in the outburst with a peak around the flux maxima. We thus test the correlation (Figure~\ref{fig:nh}) between $N_{\rm H}$ and the Eddington ratio assuming a BH mass of $10~M_\odot$, which is an average estimation above. $N_{\rm H}$ keep about $3\times 10^{22}~\rm cm^{-2}$ when unabsorbed $R_{\rm Edd}$ (upper panel of Figure~\ref{fig:nh}) is low ($< 0.1$) and start to increase as the increasing of $R_{\rm Edd}$, consistent with the scenario that outflows are driven by high accretion rate. Especially in high intermediate state, $N_{\rm H}$ is significantly larger. However, we also note that there is a degeneracy between $N_{\rm H}$ and unabsorbed $R_{\rm Edd}$, as a higher $N_{\rm H}$ will recover a larger unabsorbed $R_{\rm Edd}$. Therefore, we test the correlation between $N_{\rm H}$ and observed $R_{\rm Edd}$, and found that high observed $R_{\rm Edd}$ also correspond to large $N_{\rm H}$. Moreover, except for observations that the source are in the hard source, $N_{\rm H}$ also seems to be correlated with the unabsorbed $R_{\rm Edd}$ of disk component, and starts to enhance around several percent of $R_{\rm Edd}$.

\begin{figure*}
\includegraphics[width=0.33\textwidth]{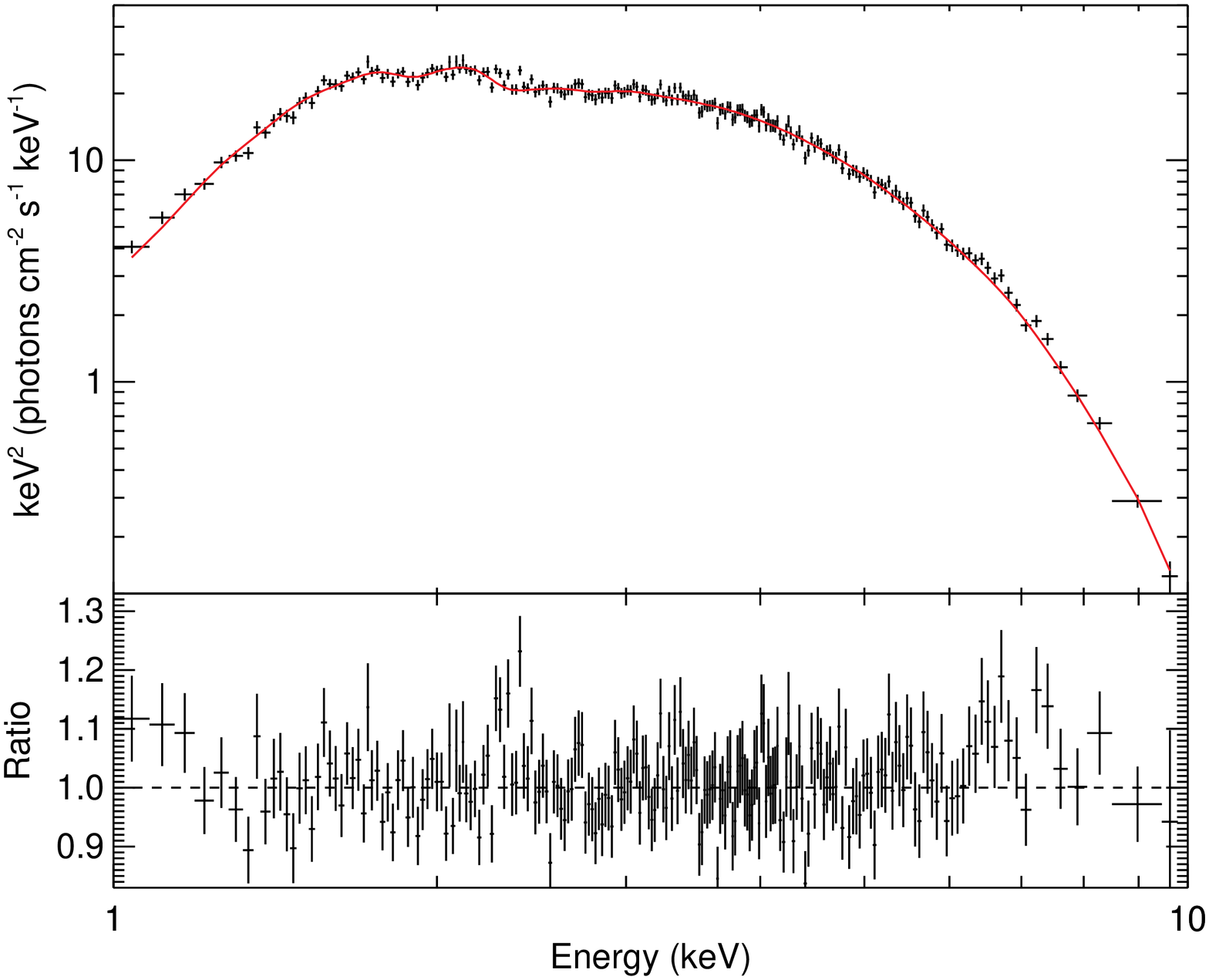} 
\includegraphics[width=0.33\textwidth]{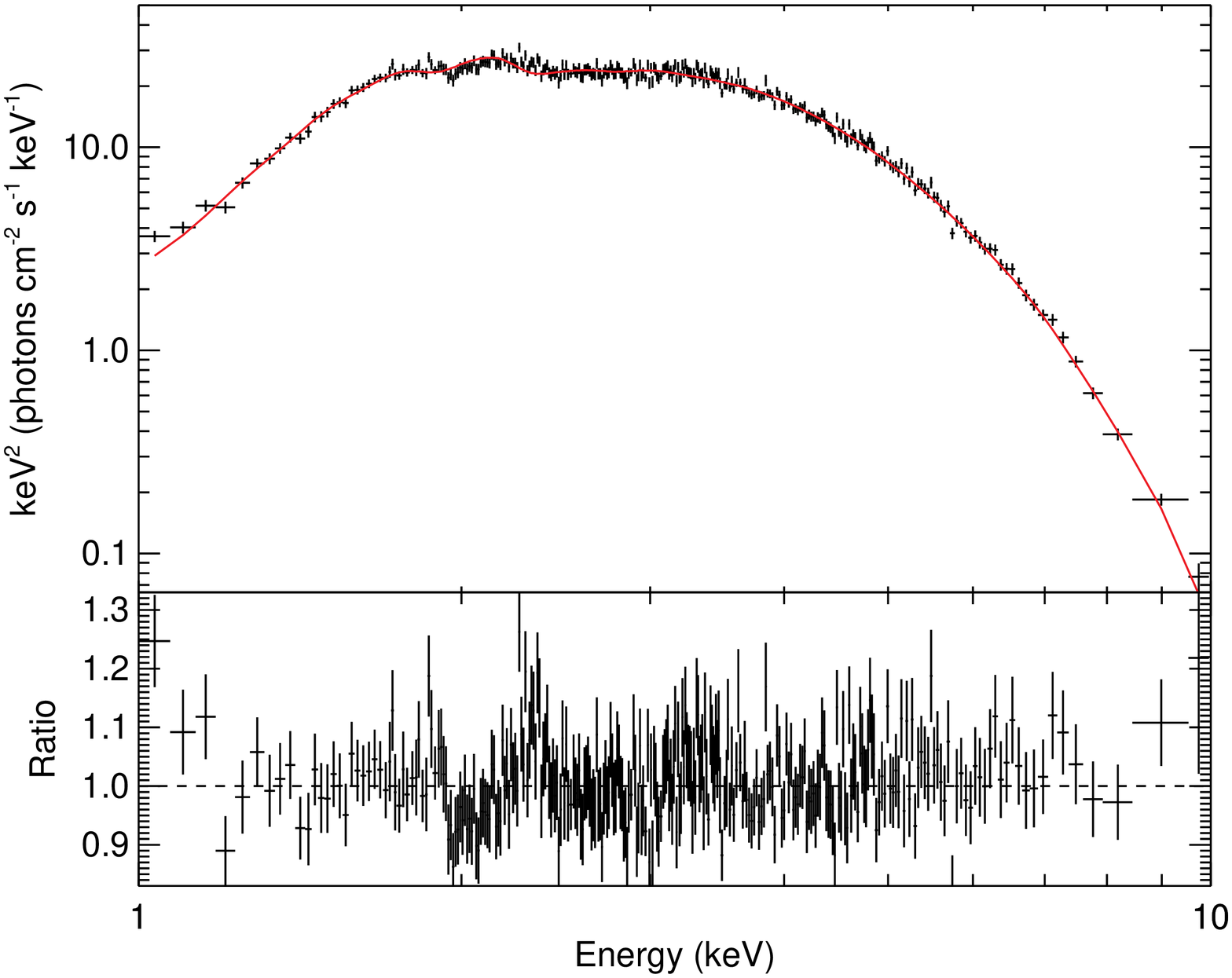}
\includegraphics[width=0.33\textwidth]{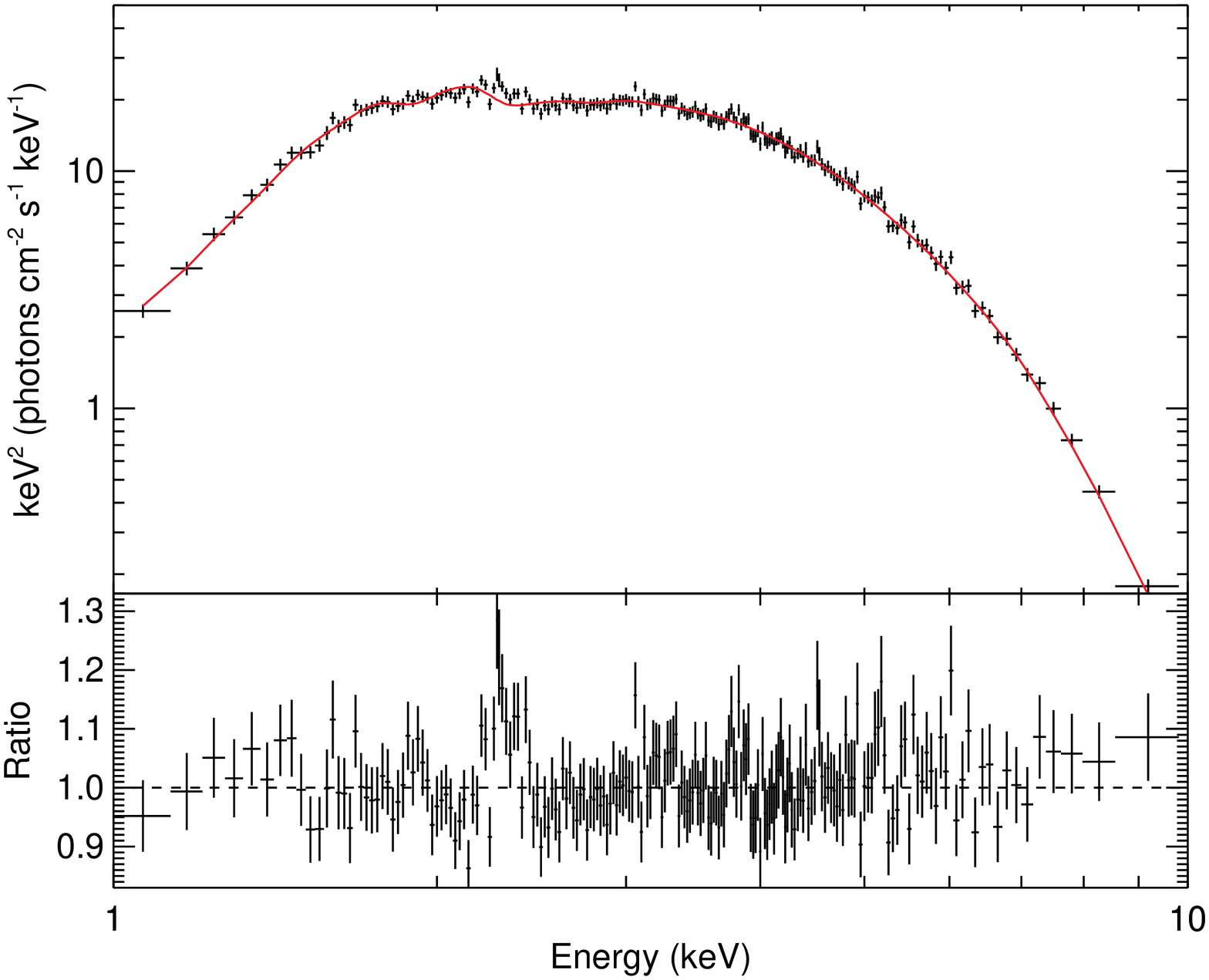}  
\caption{Examples of line-like spectral features. The data were grouped to have a signal-to-noise ratio of at least 15 per bin. Left: ObsID 10264007; middle: ObsID 102640012; right: ObsID 88245003.  
\label{fig:res}}
\end{figure*}

%line
Blueshifted X-ray absorption features are discovered in some super-Eddington sources \citep{Pinto2016, Walton2016}, and explained as originating in a wind flowing toward us. For MAXI J1535-571, some low-energy line-like residuals, such as an absorption feature at $\sim 2.0$ keV and an emission feature at $\sim 2.3$ keV, appear in the spectra (Figure~\ref{fig:res}). But the detections of the line features are marginal\footnote{The line feature is so weak that it will not affect the continuum fitting results.}, and we note that the residuals near the Silicon and Gold edges ($\sim 2$~keV) were previously observed in some XRT spectra\footnote{http://www.swift.ac.uk/analysis/xrt/digest\_cal.php\#res}. Although the improved calibration files have reduced the residuals, in a few cases, the feature is still apparent. Therefore, we do not go further in this paper based on the current data quality. High spectral resolution data are needed to further resolve the feature in MAXI~J1535-571.

%%%%%%%%%%%%%%%%%%%%%%%%%%%%%%%%%%%%%%%%%%%%%%%%%%%%%%%%%%%%%%%%%%%%%%%%%%

%%%%%%%%%%%%%%%%%%%%%%%%%%%%%%%%%%%%%%%%%%%%%%%%%%%%%%%%%%%%%%%%%%%%%%%%%%

\section*{Acknowledgments} 

We thank the anonymous referee for useful comments that have improved the paper. LT acknowledges funding support from the National Natural Science Foundation of China (NSFC) under grant numbers U1838115, the CAS Pioneer Hundred Talent Program Y8291130K2 and the Scientific and technological innovation project of IHEP Y7515570U1. CG acknowledges support from CAS President's International Fellowship Initiative (PIFI). JQ thank support from the Chinese NSFC 11673023. We thank support from National Key R\&D Program of China (grant No. 2016YFA0400800), the Chinese NSFC U1838201, 11733009 and 11473027, XTP project XDA 04060604, the Strategic Priority Research Programme 'The Emergence of Cosmological Structures' of the Chinese Academy of Sciences, Grant No.XDB09000000,and the Chinese NSFC U1838202.

\label{lastpage}
\end{document}